\documentclass[fleqn,usenatbib]{mnras}

\usepackage{newtxtext,newtxmath}
\usepackage[T1]{fontenc}

\DeclareRobustCommand{\VAN}[3]{#2}
\let\VANthebibliography\thebibliography
\def\thebibliography{\DeclareRobustCommand{\VAN}[3]{##3}\VANthebibliography}

\usepackage{graphicx}	% Including figure files
\usepackage{comment}

\newcommand{\tco}{$^{13}$CO}
\newcommand{\cdo}{C$^{18}$O}
\newcommand{\kms}{\mbox{km~s$^{-1}$}}
\newcommand{\Msun}{\mbox{$\text{M}_{\sun}$}}
\newcommand{\mm}{Mystic Mountains}

\title[ALMA obs of the Mystic Mountains]{Into the Mystic: ALMA ACA observations of the Mystic Mountains in Carina}

\author[M. Reiter et al.]{
Megan Reiter,$^{1}$\thanks{E-mail: megan.reiter@rice.edu (MR)}
P. D. Klaassen,$^{2}$
L. Moser-Fischer,$^{3}$
A. F. McLeod,$^{4,5}$
and D. Itrich$^{6}$
\\
$^{1}$Department of Physics and Astronomy, Rice University, 6100 Main St - MS 108, Houston, TX 77005, USA \\
$^{2}$UK Astronomy Technology Centre, ROE, Blackford Hill, Edinburgh, EH9 3HJ, UK \\
$^{3}$University of Bonn, Argelander-Institut fuer Astronomie, European ALMA Regional Centre -- German node, Auf dem Huegel 71, 53121 Bonn, Germany \\
$^{4}$Centre for Extragalactic Astronomy, Department of Physics, Durham University, South Road, Durham DH1 3LE, UK \\
$^{5}$Institute for Computational Cosmology, Department of Physics, University of Durham, South Road, Durham DH1 3LE, UK \\
$^{6}$European Southern Observatory, Karl-Schwarzchild-Str. 2, D-85748 Garching bei M\"{u}nchen, Germany
}

\date{Accepted 2023 August 27. Received 2023 August 21; in original form 2023 March 31}

\pubyear{2023}

\begin{document}
\label{firstpage}
\pagerange{\pageref{firstpage}--\pageref{lastpage}}
\maketitle

\begin{abstract}
We present new observations of the Mystic Mountains cloud complex in the Carina Nebula using the ALMA Atacama Compact Array (ACA) to quantify the impact of
strong UV radiation on the structure and kinematics of the gas. 
Our Band~6 observations target CO, \tco, and \cdo; we also detect DCN J=3-2 and $^{13}$CS J=5-4. 
A dendrogram analysis reveals that the \mm\ are a coherent structure, with continuous emission over $-$10.5~\kms\ $<$ v < $-$2~\kms. We perform multiple analyses to isolate non-thermal motions in the \mm\ 
including computing the turbulent driving parameter, $b$, which indicates whether compressive or solenoidal modes dominate.
Each analysis yields values similar to other pillars in Carina that have been observed in a similar way but are subject to an order of magnitude less intense ionizing radiation. We find no clear correlation between the velocity or turbulent structure of the gas and the incident radiation, in contrast to other studies targeting different regions of Carina. 
This may reflect differences in the initial densities of regions that go on to collapse into pillars and those that still look like clouds or walls in the present day. Pre-existing over-densities that enable pillar formation may also explain why star formation in the pillars appears more evolved (from the presence of jets) than in other heavily-irradiated but non-pillar-like regions. High resolution observations of regions subject to an array of incident radiation are required to test this hypothesis. 
\end{abstract}


%%%%%%%%%%%%%%%%%%%%%%%%%%%%%%%%%%%%%%%%%%%%%%%%%%

%%%%%%%%%%%%%%%%% BODY OF PAPER %%%%%%%%%%%%%%%%%%

\section{Introduction}

Feedback is the principal process by which molecular clouds are destroyed \citep{matzner2002}. 
High-mass stars inject energy and momentum into their surroundings via winds and radiation before their eventual deaths as supernovae. 
Growing evidence suggests that pre-supernova feedback plays a dominant role reshaping the interstellar medium (ISM) and regulating star formation \citep[e.g.,][]{kruijssen2019,mcleod2021,chevance2022}. 
Feedback has also been invoked as the stimulus for the formation of new stars, either by stimulating 
the collapse of existing cores or collecting material into new ones \citep[e.g.,][]{bertoldi1989,elmegreen1977}.

Many theoretical models have been developed to address the question of how stellar feedback
shapes the surrounding gas. Models can produce dust pillars like those seen in many H~{\sc ii} regions
\citep[e.g.,][]{hester1996} either through growing instabilities or by revealing pre-existing substructure
(e.g., filaments and cores) as the surrounding, lower-density material is more easily swept away 
\citep[][]{gritschneder2010,dale2012,tremblin2012_sc,tremblin2012_sct,walch2013,menon2020}. Despite their morphological similarities, models predict
differences in the following: density contrasts between the pillars and the surrounding medium, progression speeds
of the ionization front, star formation efficiencies, and timescales, particularly whether cores
are already present or formed by the ionization-driven shock. This results in measurable differences in the gas kinematics within (and around) the photodissociation region (PDR) interface.

Despite the diagnostic potential of cold gas kinematics, only a few studies exist with high spatial resolution to measure the variation of the cold gas kinematics in dust pillars in star-forming regions.  
Far-IR observations from SOFIA and \emph{Herschel} provide kinematic evidence for dust pillars forming via collapse  on the edges of H~{\sc ii} regions and small globules produced by turbulent fragmentation 
\citep{schneider2012,tremblin2013}. 
However, it is only with millimeter interferometry that dust pillars can be spatially and spectrally resolved.
One of the first such studies was \citet{klaassen2014} who mapped a pillar in Vulpecula. 
The observed pillar properties are most consistent with models that have low velocity dispersions, but no one model matches all of the observed gas kinematics \citep[e.g.,][]{gritschneder2009,gritschneder2010,dale2012}.

To probe a broader range of conditions, \citet{klaassen2020} presented a survey of 13 dust pillars in the Carina Nebula that sample a range of morphologies and environments, including the incident ionizing flux. 
Many of these reside in the actively star-forming South Pillars where star formation may have been 
triggered by the hundreds of 
O- and B-type stars in the region \citep{smith2010_spitzer,berlanas2023}. 
Cold gas kinematics are broadly consistent with pillars forming from turbulent media as they are sculpted by ionizing radiation. 
However, this sample does not include the most intense ionizing radiation that affects the gas nearest the central clusters of Carina.

\begin{figure}
	\includegraphics[width=\columnwidth]
{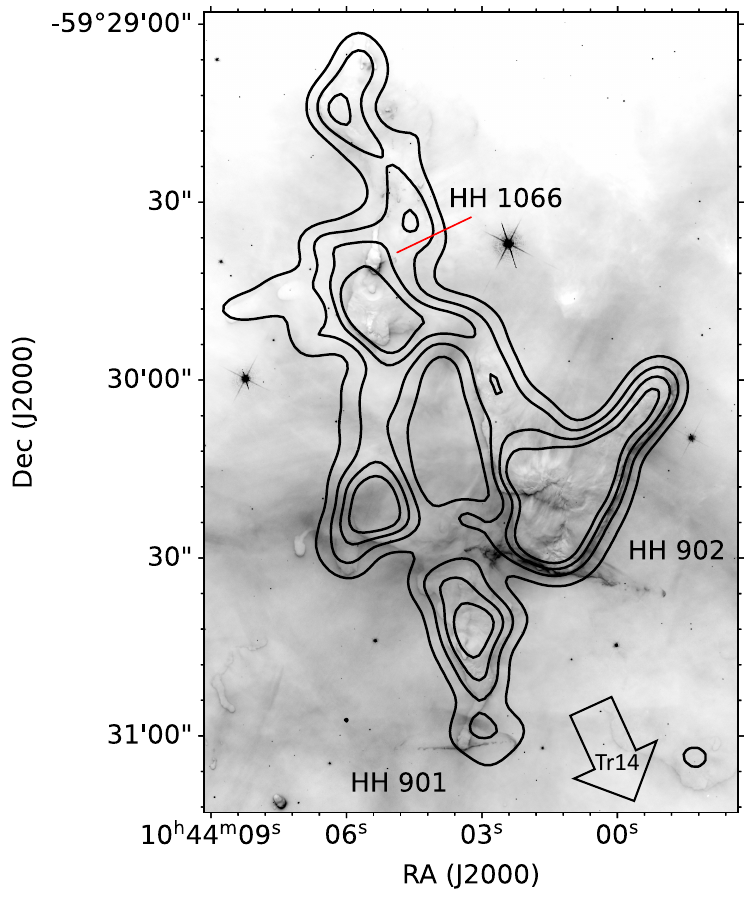}
    \caption{CO J=2-1 contours on an H$\alpha$ image from \emph{HST}. The famous jets, HH~901, HH~902, and HH~1066, are labeled. 
    The CO emission is integrated over the velocity range $-10.5 < v < -2$~\kms; contours are 20, 40, 60, 80, and 100\% of the peak intensity. } 
    \label{fig:hst_contours}
\end{figure}

To probe the most intense feedback in the Carina Nebula, we target a cloud complex that is heavily irradiated by the young, massive cluster Trumpler 14 (Tr14). 
The so-called \mm\footnote{as the region was dubbed when imaged to commemorate the 20$^{\mathrm{th}}$ anniversary of the \emph{Hubble Space Telescope}} \citep[Area~29 in][]{har15} lie 
$\sim$1~pc to the north of Tr14. 
Copious ionizing photons ($Q_H \sim 10^{50}$~s$^{-1}$; \citealt{smith2006_energy}) 
illuminate and sculpt multiple pillars in the \mm.  
Three famous Herbig-Haro (HH) jets emerge from the tips of these pillars -- HH~901, HH~902, and HH~1066  \citep[see Figure~\ref{fig:hst_contours} and][]{smith2010,reiter2013,reiter2014}. 
However, the pillars themselves are largely opaque and only a few protostars are detected in the infrared \citep[IR; e.g.,][]{povich2011,ohl12}.  
With modest angular resolution ($\sim 2\arcsec$), only the HH~1066 jet-driving source was directly detected in the infrared  \citep{ohl12,reiter2016,reiter2017}. 
The HH~1066 driving source is also one of only two sources in Carina with a marginally resolved circumstellar disk \citep{mesa-delgado2016}. 
\citet{reiter2017} argued that more intense feedback closer to Tr14 may have compressed the gas, leading to high densities that obscure the HH~901 and HH~902 jet-driving sources in the IR. 
Indeed, the first detections of the HH~901 and HH~902 driving sources were only recently reported by \citet{cortes-rangel2020}.

In this paper, we present ALMA observations using the Atacama Compact Array (ACA; also known as the Morita Array) of the entire Mystic Mountains complex. 
The ionizing photon flux incident on the Mystic Mountains is an order of magnitude higher than the pillars in 
\citet{klaassen2020}. 
The complex is large ($\sim 1\arcmin\ \times 2\arcmin$), and
thus samples a range of incident ionizing flux 
within a single pillar complex. 
By studying gas
kinematics in the Mystic Mountain, we will constrain the role of ionizing radiation in stimulating or
starving future star formation, a key test of the role of feedback in regulating star formation.

%

%=======================================================================
\section{Observations}
%=======================================================================
%
\begin{figure*}
    \centering
    \includegraphics[scale=0.35]{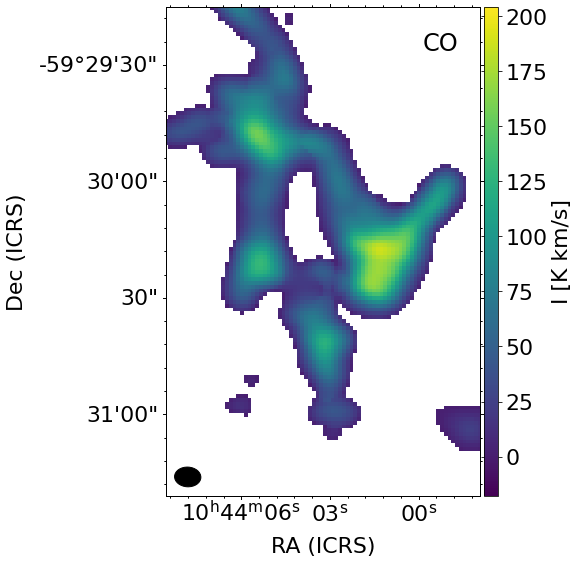}
    \includegraphics[scale=0.35]{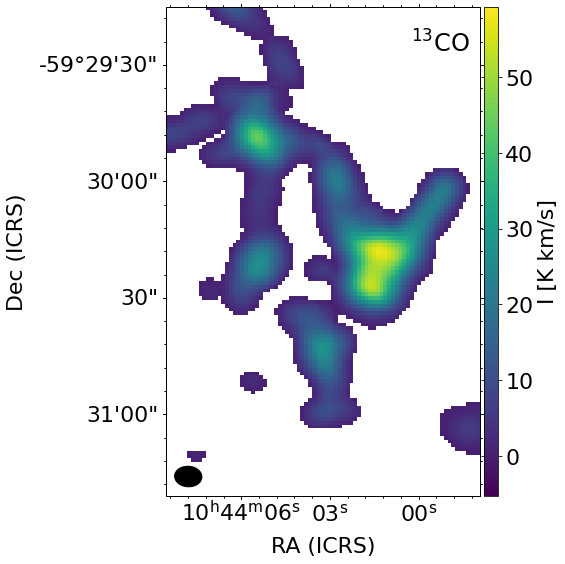}
    \includegraphics[scale=0.35]{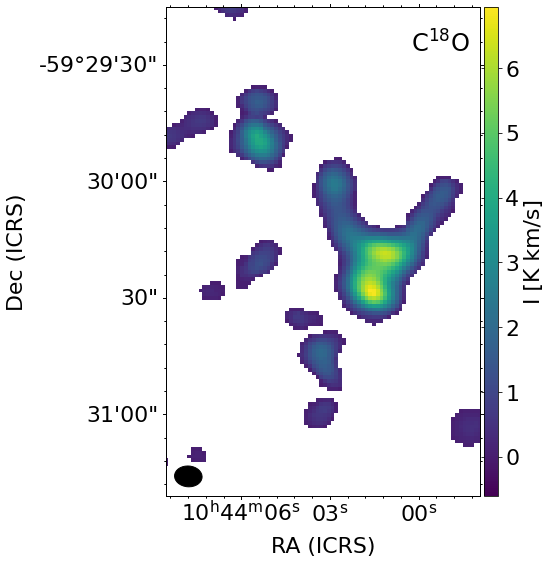}
    \includegraphics[scale=0.35]{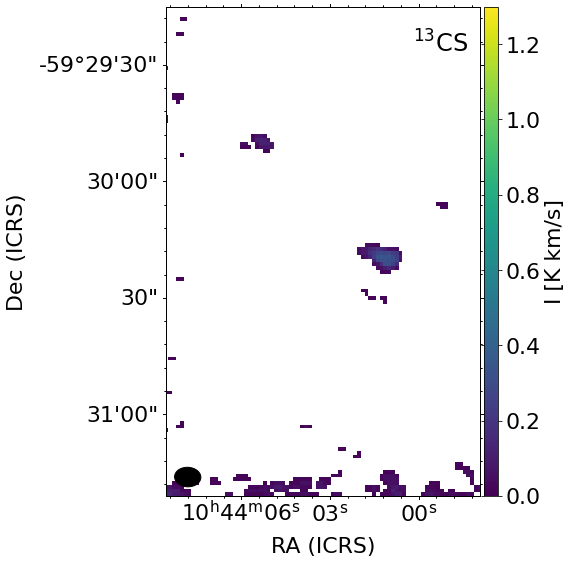}
    \includegraphics[scale=0.35]{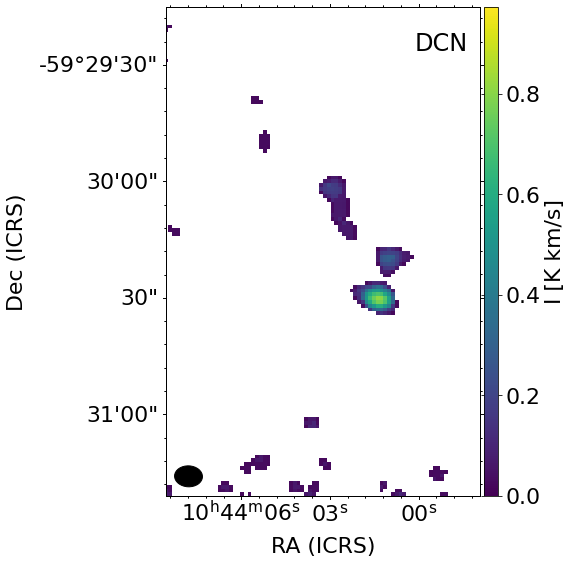}
    \caption{Integrated intensity (moment~0) maps of the lines detected in this study -- CO, \tco, \cdo, $^{13}$CS, and DCN. All CO isotopologues are detected with $>5\sigma$ significance; DCN and $^{13}$CS are detected with $>3\sigma$ significance. }
    \label{fig:mom0_maps}
\end{figure*}

ALMA ACA Band 6 observations of the \mm\  
were obtained on 19 January 2019. 
Observations are a mosaic of 21 pointings of the 7m-array (12 antennae) 
mosaic centre 
R.A.=10\fh44\fm02\fs010, 
%10:44:02.0417
decl.=$-$59$^\circ$30\arcmin01\farcs0 (ICRS). 
The maximum recoverable scale (MRS) is
28.6\arcsec. 
We also obtained Total Power (TP) data to ensure our observations capture emission from the largest scales. 
These were obtained on 28-29 November 2018 with a third epoch on 04 December 2018. 

Our observational setup targeted rotational transitions J=2-1 of the CO isotopologues $^{12}$CO, $^{13}$CO and C$^{18}$O, as well as SiO J=5-4 and $^{13}$CS J=5-4. 
All observations were imaged to a velocity resolution 0.17 \kms.  
We also observed a continuum spectral window with resolution $\sim 40$ \kms. 
When combined with line-free channels in other bands, continuum emission covers approximately 2.5~GHz in Band 6. 

Bandpass, flux, and gain calibration were done with external calibrators using the \emph{Common Astronomy and Software Applications} \citep[CASA,][]{mcm07} v5.4.0-70. 
The bandpass and flux calibrators for the Band 6 observations performed in January 2019 were J0940$-$6107 and J1047$-$6217.

Continuum subtraction and cleaning via the imaging pipeline yielded insufficient results for our needs, 
so continuum ranges were identified by eye and a deeper clean performed using 
\texttt{tclean} in CASA v6.4.4 to produce images. 
We applied auto-masking and Briggs-weighting with a robust parameter of 0.5. We deconvolved the image in multiscale mode with scale of 0, 5, and 15 times the pixel size of 1.0 arcsec.
The absolute flux scaling uncertainty is estimated to be about 15\%.
The synthesized beamsizes of the reduced data  
range typically between 5\farcs4--7\farcs2, 
corresponding to a spatial resolution 
12,420--16,560~AU at the distance of Carina \citep[2.3~kpc;][]{smith2006_distance,goppl2022}.

Finally, the ACA and TP data were combined using the \texttt{CASA} task \texttt{feather}. 
This task regrids the lower resolution data to match the higher resolution data, scales them by the ratio of the clean beams, then combines the two datasets in fourier space before transforming back to the image plane. 
This provides much better recovery of extended emission than the ACA alone, allowing us to capture structures up to $\sim$27$''$ at 230 GHz \citep[see][for an example]{klaassen2020}.

%%=============================================================================

\begin{table*}
\caption{Spectral and imaging characteristics of the data.}
\begin{center}
\begin{footnotesize}
\begin{tabular}{lllllllll}
\hline\hline
Name & Frequency & Bandwidth & Resolution & $\theta_{\rm min}$ & $\theta_{\rm max}$ & P.A. & RMS & Comment \\ 
& [GHz] & [MHz] & [km/s] & [\arcsec] & [\arcsec] & [$^\circ$] & [mJy bm$^{-1}$]\\
\hline\hline
\multicolumn{9}{c}{Molecular lines} \\
\hline
SiO J=5-4       & 217.1049800   & 468.75    & 0.337 &  5.39 & 7.18 & 87.6 & 100.3 & not detected \\ 
DCN J=3-2 & 217.23855   & 468.75    & 0.337 &  5.39 & 7.18 & 87.6 & 100.3 & in SiO spectral window \\ 
\cdo\ J=2-1 & 219.5603568   & 117.19    & 0.167 &  5.35 & 7.09 & 85.9 & 141.8 & \\ 
\tco\ J=2-1 & 220.3986765   & 117.19    & 0.167 &  5.30 & 7.04 & 86.9 & 245.4 &\\ 
$^{12}$CO J=2-1 & 230.538       & 117.19   & 0.159 &  5.07 & 6.77 & 87.0 & 993.0 &\\ 
$^{13}$CS J=5-4        & 231.220686   & 117.19    & 0.158  & 5.02 & 6.77 & 87.5 & 242.3  &\\
\hline
\multicolumn{9}{c}{Continuum} \\
\hline
B6          & 225.8799    & 2530.0    & 40.038  &  4.95 & 6.75 & 86.9 & 1.700$^*$& \\ 
\hline 
\multicolumn{9}{l}{$^*$ RMS of the aggregated bandwidth image. } \\
\end{tabular} 
\end{footnotesize}
\end{center}
\label{t:ALMA_lines}
\end{table*}

\subsection{Complementary data from the \emph{Hubble Space Telescope (HST)}} 
We compare the ALMA observations to a narrowband H$\alpha$ image from \emph{HST}. 
A four-point mosaic of the field containing HH~901, HH~902, and HH~1066, dubbed "The Mystic Mountains," was taken February 1-2, 2010 to commemorate the 20th anniversary of \emph{HST} (PID~12050, P.I. M.\ Livio). 
Images were obtained with the UVIS channel of the Wide Field Camera 3 (WFC3). 
The total integration time in the F657N filter was 1980~s. 
The observations and their analysis are presented in more detail in \citet{reiter2013}.

%=======================================================================
\section{Results and Analysis}\label{s:results}
%=======================================================================

Figure~\ref{fig:mom0_maps} shows the integrated intensity (moment~0) maps of all  emission lines detected in this study.  
Extended CO emission traces the kinematics and cloud structure throughout the Mystic Mountains complex with knots of bright emission in \tco\ and \cdo\ tracing $\sim 1-2$ clumps of emission in each pillar. 
The most complex emission is in the pillar with the HH~902 jet. 
Two emission peaks in the CO and isotopologues hint at multiple star-forming clumps.

Visual inspection of the CO datacube reveals that emission associated with the Mystic Mountains is contained within the velocity range $-15 < v < 5$~\kms. 
This range includes the systemic velocities of the HH~901 ($-5.0$~\kms) and HH~902 ($-8.5$~\kms) pillars identified by \citet{cortes-rangel2020}. 
Pillar-like emission to the east and north of these two pillars contains HH~1066 (for which we estimate a systemic velocity of $-6.5$~\kms). 
Additional CO emission not associated with the \mm\ is detected at other velocities ($\pm 20$~\kms\ from the $v_{\mathrm{LSR}}$ of Carina, $-20$~\kms), but we do not discuss this further. 

To determine the precise velocity range and extent of the gas associated with the Mystic Mountains complex, we use a dendrogram analysis to identify coherent structures in position-position-velocity space. 
As described in \citet{rosolowsky2008} and \citet{goodman2009}, dendrograms provide a hierarchical representation of data, aiding the analysis of physical conditions on multiple scales from a single dataset. 
Large coherent features that are not part of another larger, parent structure are `trunks'; these contain substructures called `branches' that are composed of individual local maxima, or `leaves,' that cannot be subdivided further.

We use the python packages {\sc astrodendro} \citep{robitaille2019} and {\sc scimes} \citep{colombo2015} to decompose the structures in the CO datacube. 
We use the following parameters for the decomposition: 
a minimum value (min\_value) that defines the noise threshold -- we adopt 6$\sigma$; 
a minimum intensity (min\_delta) for that defines the threshold the peak flux must exceed to be identified as a separate structure -- we adopt 2$\sigma$; and 
a minimum number of pixels for a leaf to be an independent entity -- we use the number equivalent to three beams (124 pixels in our case).

\begin{figure*}
	\includegraphics[width=\columnwidth]{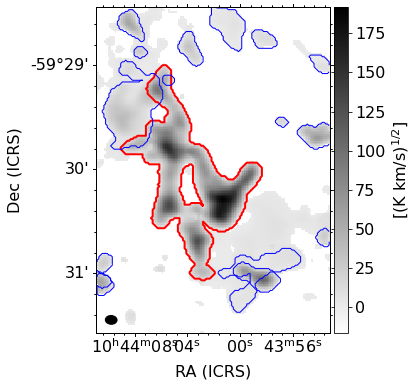}
	\includegraphics[trim=0mm -25mm 0mm 0mm,width=\columnwidth]{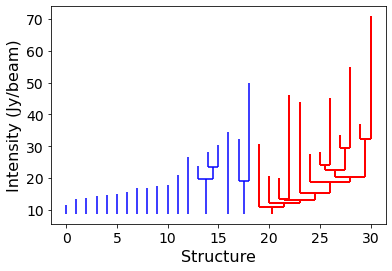}
    \caption{Applying a dendrogram analysis to the CO velocity cube ($-15 < v < 5$~km~s$^{-1}$) shows the Mystic Mountains are a single, coherent structure. 
    \textbf{Left:} Shows the CO integrated intensity with colored contours indicating dendrogram trunks. The \mm\ is shown in red; all other features are shown in blue.
    \textbf{Right:} The dendrogram tree, using the same color scheme as the left panel. The \mm, shown in red, is a separate feature with all substructures stemming from the same trunk.  
    }
    \label{fig:dendros}
\end{figure*}

\begin{figure}
	\includegraphics[width=\columnwidth]{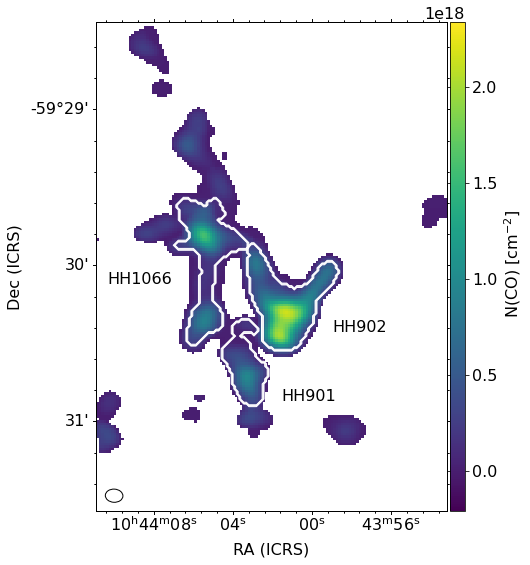}
    \caption{
    Column density map with a single white contour that outlines the branches identified by the CO dendrogram analysis. Peak and median values reported in Table~\ref{t:molecular_props} are computed within these pillar boundaries.  }
    \label{fig:pillar_boundaries}
\end{figure}

With these parameters, a dendrogram analysis identifies the Mystic Mountains as a trunk, indicating that it is a coherent cloud complex with contiguous emission over the velocity range $-10.5 < v < -2$~\kms. Figure~\ref{fig:dendros} shows the \mm\ as a single tree in red with other structures shown in blue. 
Within the \mm\ complex, the dendrogram analysis identifies three separate pillars as branches. 
Contours defining the outlines of these branches are shown in Figure~\ref{fig:pillar_boundaries}. 
We refer to the pillars by the name of the prominent jets that they host -- HH~901, HH~902, and HH~1066.

In the following sections, we use this velocity range to derive the spatially-resolved physical parameters of the cold molecular gas in the Mystic Mountains. 
We provide maps of these spatially-resolved quantities in Appendix~\ref{appendix:tau_maps} and \ref{appendix:Ncol_maps}. 
To provide representative values in Table~\ref{t:molecular_props}, we compute the values within the dendrogram branches that define each of the pillars. 

%

%%======================================================================
\begin{table*}
\caption{Summary of molecular line derived physical properties. Columns are the species/transition, peak and median intensities, median column density if optically thin, median optical depth, and median column if optically thick, respectively. Spatially resolved maps of these quantities are shown in Appendices~\ref{appendix:tau_maps}--\ref{appendix:Ncol_maps}. \label{t:molecular_props}}
\centering
\begin{footnotesize}
\begin{tabular}{rlllllll}
\hline\hline
Line & $I_{\mathrm{peak}}$ &
$I_{\mathrm{median}}$ & 
 log(N$_{\mathrm{thin}}$)$_{\mathrm{max}}$ & 
 log(N$_{\mathrm{thin}}$)$_{\mathrm{median}}$ & $\tau_{\mathrm{max}}$ &  log(N$_{\mathrm{thick}}$)$_{\mathrm{peak}}$ & 
 log(N$_{\mathrm{thick}}$)$_{\mathrm{median}}$\\ 
        & [K~km~s$^{-1}$] & [K~km~s$^{-1}$] & 
         [cm$^{-2}$] & [cm$^{-2}$] & 
         & [cm$^{-2}$] & [cm$^{-2}$] \\
\hline
\multicolumn{8}{c}{ \mm; $v_{\mathrm{LSR}} \approx -6.7$~\kms$^{\ddagger}$ } \\ 
\hline
$^{12}$CO J=2-1 & 185.0 & 46.5 & 17.0 & 16.2 & 44.6 & 18.3 & 17.2 \\ 
\tco\ J=2-1 & 57.5 & 8.8 & 16.5 & 15.4 & 4.2 & 16.6 & 15.5 \\
\cdo\ J=2-1 & 6.9 & 1.14 & 15.6 & 14.8 & \ldots & \ldots & \ldots \\
\hline
\multicolumn{8}{c}{HH~901 pillar; $v_{\mathrm{LSR}} \approx -5.0$~\kms$^{\dagger}$} \\ 
\hline
$^{12}$CO J=2-1 & 114.2 & 38.8 & 16.8 & 16.4 & 33.0 & 18.0 & 17.5 \\ 
\tco\ J=2-1 & 27.0 & 9.0 & 16.1 & 15.7 & 0.68 & 16.1 & 15.7 \\
\cdo\ J=2-1 & 1.97 & 0.61 & 15.0 & 14.5 & \ldots & \ldots & \ldots \\
\hline
\multicolumn{8}{c}{HH~902 pillar; $v_{\mathrm{LSR}} \approx -8.5$~\kms$^{\dagger}$} \\ 
\hline
$^{12}$CO J=2-1 & 174.1 & 65.6 & 17.0 & 16.6 & 44.6 & 18.3 & 17.9 \\
\tco\ J=2-1 & 56.8 & 18.3 & 16.5 & 16.0 & 1.9 & 16.6 & 16.0 \\
\cdo\ J=2-1 & 6.90 & 1.84 & 15.6 & 15.0 & \ldots & \ldots & \ldots\\
\hline
\multicolumn{8}{c}{HH~1066 pillar; $v_{\mathrm{LSR}} \approx -6.5$~\kms} \\ 
\hline
$^{12}$CO J=2-1 & 138.9 & 42.0 & 16.9 & 16.6 & 35.0 & 18.2 & 17.6 \\
\tco\ J=2-1 & 42.1 & 7.65 & 16.4 & 15.7 & 3.1 & 16.4 & 15.8 \\
\cdo\ J=2-1 & 4.03 & 0.81 & 15.3 & 14.7 & \ldots & \ldots & \ldots\\
\hline
\multicolumn{8}{l}{$^{\dagger}$ $v_{\mathrm{LSR}}$ from \citet{cortes-rangel2020} }\\
\multicolumn{8}{l}{$^{\ddagger}$ $v_{\mathrm{LSR}}$ from the intensity-weighted average velocity of the \mm\  }\\
\end{tabular}
\end{footnotesize}
\end{table*}
%
%%==========================================================================

\subsection{Optical depth}

We compute the optical depth at each position and velocity where emission is detected with a significance $\geq 5\sigma$ using the following expression \citep[equation~1 from][]{choi1993}: 
\begin{equation}
\frac{T_{\rm main,v}}{T_{\rm iso,v}} = \frac{1 - e^{-\tau_{\rm main,v}}}{1-e^{- \tau_{\rm iso,v}}} = \frac{1-e^{-\tau_{\rm main,v}}}{1-e^{- \tau_{\rm main,v}/R}}
\end{equation}
where ``main'' is the more abundant species and ``iso'' is the optically thin (isotopologue) transition used to correct it. 
$R$ is the scale factor for the relative abundance of the two species. 
We use [$^{12}$CO/\tco]$=60$ (\citealt{rebolledo2016}, see also \citealt{jacob2020})
and [$^{12}$CO/\cdo]$=560$ \citep{wilson1994}. 
We assume the same excitation temperature for both molecules. 
We find that CO emission is optically thick over a large portion of the \mm\ whereas \tco\ is only optically thick near the brightest emission in the HH~902~pillar. 
Maps of the spatially-resolved optical depth at the source velocity are shown in Appendix~\ref{appendix:tau_maps} and maximum values are reported in Table~\ref{t:molecular_props}.

\subsection{Molecular column density}\label{ss:alma_NH}
We compute the column density of each observed transition using the following equation \citep[see e.g.,][]{mangum2015}: 
\begin{equation}
  N_{tot} = \frac {8\pi k \nu^2Q(T_{ex}) e^{E_{u} / kT_{ex}} J_{\nu}(T_{ex}) }
    {h c^3 g_u A_{ul} [J_{\nu}(T_{ex}) - J_{\nu}(T_{\rm cmb})] } 
    \int 
    T_{\rm mb} \frac{ \tau_{v}}{1 - e^{-\tau_{v}}} dv  \,\,\mathrm{cm}^{-2}
\end{equation}\label{eq:Ncol}
where
$Q(T_{ex})$ is the rotational partition function for a given excitation temperature,  
$g_u$ is the rotational degeneracy of the upper level with energy $E_u$, 
$A_{ul}$ is the Einstein A coefficient for the transition, 
$k$ is the Boltzmann constant, 
$h$ is the Planck constant, 
%$B_{\nu}$ is the Planck function,
$J_{\nu} =(h\nu/k)/[\exp(h\nu/kT)-1]$ is the Planck function in temperature units (K), and 
$\tau_{v} / (1 - e^{-\tau_{v}})$ is a correction factor for non-zero optical depth \citep[see, e.g.,][]{goldsmith1999}. 
This assumes that all transitions have the same T$_{\mathrm{ex}}$.
Physical parameters for the relevant molecules and transitions (frequency, rotational partition function, Einstein A coefficients, etc.) were obtained from the JPL Spectral Line Catalog \citep{pickett1998} and the Leiden Atomic and Molecular Database \citep[LAMBDA;][]{schoier2005}. 

To compute the column density, we assume an excitation temperature $T_{\mathrm{ex}}=30$~K. 
We assume that the gas temperature is the same as the dust temperature derived from the far-IR spectral energy distribution (SED) that \citet{roccatagliata2013} used to compute a temperature map of the entire Carina region.
As discussed in \citet{mangum2015}, assuming a single temperature is often a poor assumption and we expect that to be true for the \mm.  
However, adopting higher excitation temperatures ($T_{ex}\approx40-80$~K) 
or a variable excitation temperature within this range changes the estimated column density by a factor of $\lesssim 2$.  
In the absence of a better temperature measurement, we adopt a single number in this study. 
Maps of the spatially-resolved column density calculation for each of the CO isotopologues are shown in Appendix~\ref{appendix:Ncol_maps} and median column densities in each pillar are reported in Table~\ref{t:molecular_props}.

\subsection{Molecular gas mass of the \mm}\label{ss:alma_masses}

We estimate the mass of the \mm\ complex from the optical-depth corrected spatially-resolved 2D CO column density map using the equation: 
\begin{equation}
M_{gas} = \mu_g m({\rm H_2}) A \left[ \frac{\rm H_2}{\rm CO} \right] \Sigma N({\rm CO}) 
\end{equation}
where $\left[\mathrm{H_2}/\mathrm{CO} \right] = 1.1 \times 10^4$ \citep{pineda2010} is the abundance of H$_2$ compared to CO, 
$\mu_g = 1.41$ is the mean molecular weight \citep{kauffmann2008}, 
$m(H_2)$ is the mass of molecular hydrogen, and 
$A$ is the area of each pixel in the map. 
We compute a mass of $\sim$36~\Msun\ for the entire Mystic Mountains complex. 
More recent measurements find $\left[\mathrm{H_2}/\mathrm{CO} \right] = 6000$ \citep{lacy2017} which reduce the estimated mass by a factor of 2.

\subsection{Clumps, cores, and YSOs}\label{ss:clumps_cores}

\subsubsection{\cdo\ clumps}\label{ss:virial_masses}
\begin{figure*}
	\includegraphics[width=0.965\columnwidth]{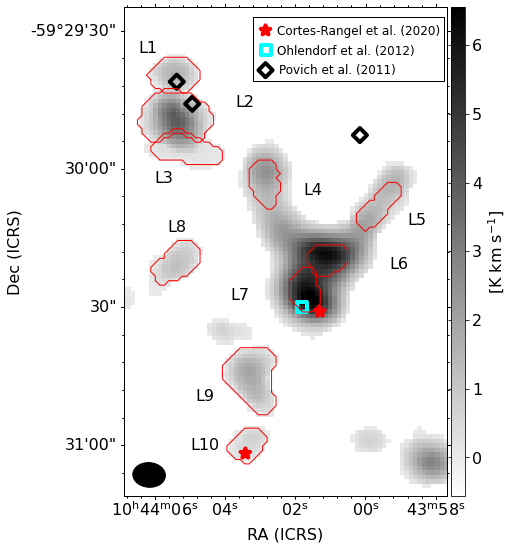}
	\includegraphics[width=\columnwidth]{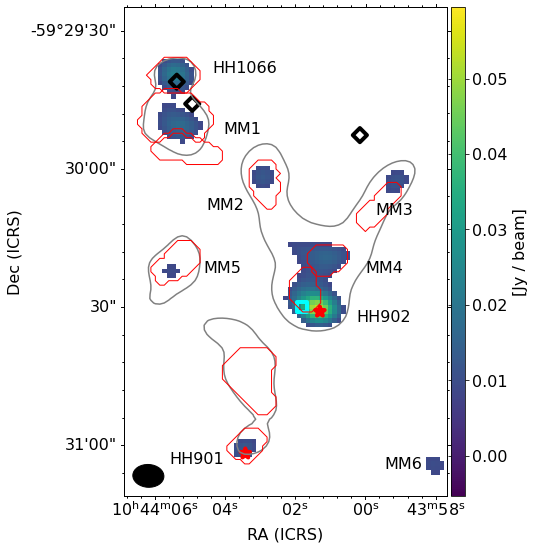}
	\caption{
	\textbf{Left:} Local maxima (leaves) identified by the dendrogram analysis shown in red contours on a moment~0 map of \cdo. 
	\textbf{Right:} Dendrogram leaves (red contours) compared to continuum peaks shown in the colorscale. A single gray contour shows the \cdo\ emission at 20\% of the peak emission. 
	Both panels show YSOs detected in other surveys: 
    black diamonds are YSOs from \citet{povich2011}; 
    the cyan square is the point source from \citet{ohl12}; and the red stars are the YSOs identified by \citet{cortes-rangel2020}. 
	}
    \label{fig:dendro_leaves}
\end{figure*}
%
%%----------------------------------------------------------------
\begin{table*}
\caption{Physical properties of dendrogram leaves shown in Figure~\ref{fig:dendro_leaves}.\label{t:leaves}}
%\centering
\begin{footnotesize}
\begin{tabular}{llcccccccccl}
\hline\hline
Leaf & MM & R.A. & decl. & v$_{\mathrm{src}}$& $<R>$ & $\Delta v$ & $\sigma_{\mathrm{turb;1D}}$ & M$_{H_2}$ & M$_{vir}$ & stable? & comment \\
        & & (J2000) & (J2000) & [\kms] & [pc] & [\kms] & [\kms] & [\Msun] & [\Msun] & & \\
        \hline
L1 & HH~1066~MM & 10:44:05.4 & $-$59:29:40 & -8.4 & 0.02 & 1.3 & 0.53 & 4.9 & 4.1 & N & \\ 
L2 & MM1 & 10:44:05.4 & $-$59:29:50 & -7.1 & 0.03 & 1.5 & 0.64 & 48.7 & 9.0 & N & $^{13}$CS \\ 
L3 & & 10:44:05.2 & $-$59:29:56 & -7.0 & 0.02 & 1.3 & 0.52 & 5.6 & 4.5 & N & \\ 
L4 & MM2 & 10:44:02.9 & $-$59:30:02 & -7.5 & 0.02 & 0.93 & 0.37 & 5.9 & 1.9 & N & \\ 
L5 & MM3 & 10:43:59.7 & $-$59:30:09 & -7.0 & 0.02 & 1.1 & 0.43 & 4.5 & 2.6 & N & \\ 
L6 & MM4 & 10:44:01.1 & $-$59:30:19 & -8.0 & 0.02 & 1.1 & 0.45 & 15.7 & 2.6 & N & DCN, $^{13}$CS \\ 
L7 & HH~902~MM & 10:44:01.7 & $-$59:30:27 & -7.8 & 0.02 & 1.2 & 0.50 & 17.8 & 3.3 & N & DCN, $\sim$2.7\arcsec\ offset \\ 
L8 & MM5 & 10:44:05.3 & $-$59:30:20 & -5.1 & 0.01 & 1.1 & 0.46 & 3.5 & 3.0 & N & \\ 
L9 & & 10:44:03.3 & $-$59:30:45 & -5.1 & 0.03 & 1.1 & 0.47 & 19.7 & 4.4 & N & \\ 
L10 & HH~901~MM & 10:44:03.3 & $-$59:30:60 & -5.2 & 0.02 & 0.58 & 0.21 & 0.95 & 0.66 & N & \\ 
\hline
\end{tabular}
\end{footnotesize}
\end{table*}
%%----------------------------------------------------------------
%

We repeat the dendrogram analysis 
on the \cdo\ data to identify clumps using this higher density tracer. 
As for the CO analysis, we use the following thresholds: 
an intensity minimum value of 6$\sigma$ to ensure that we are using only well-detected emission, 
a minimum intensity of 2$\sigma$ to define a separate peak, and 
a minimum number of pixels equivalent to three beams. 
The leaves (local maxima) 
detected with this analysis are shown in Figure~\ref{fig:dendro_leaves}.

\begin{figure}
	\includegraphics[scale=0.285]{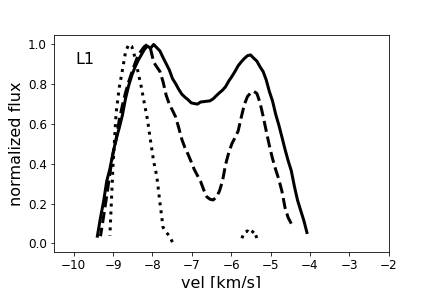}
	\includegraphics[scale=0.285]{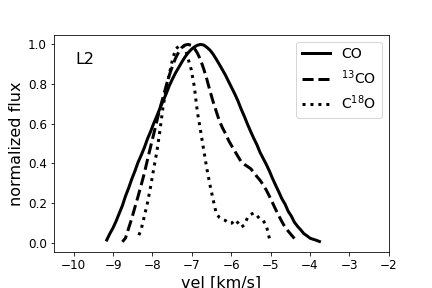}
	\includegraphics[scale=0.285]{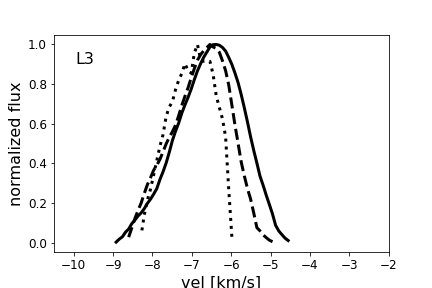}
	\includegraphics[scale=0.285]{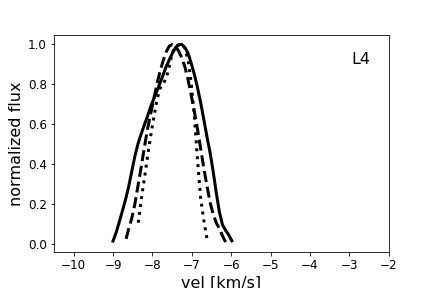}
	\includegraphics[scale=0.285]{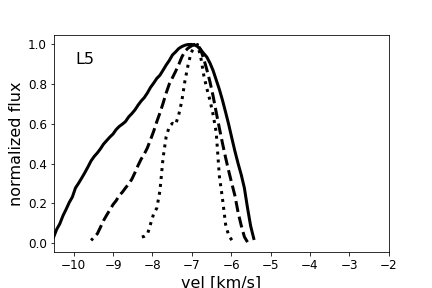}
	\includegraphics[scale=0.285]{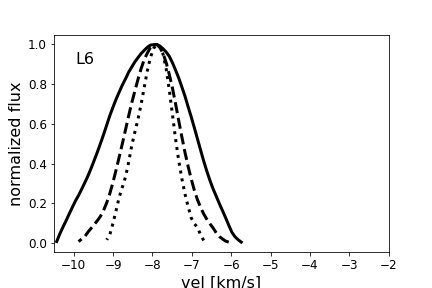}
	\includegraphics[scale=0.285]{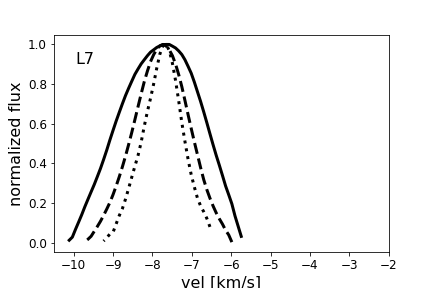}
	\includegraphics[scale=0.285]{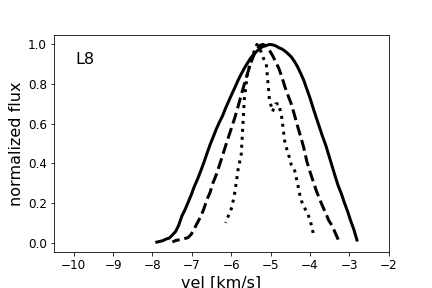}
	\includegraphics[scale=0.285]{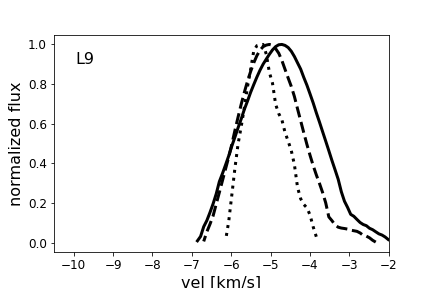}
	\includegraphics[scale=0.285]{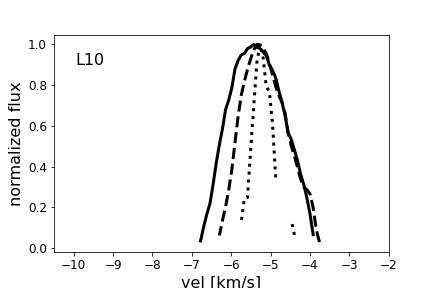}
	\caption{
	The summed intensity profile of each dendrogram leaf identified in the \cdo\ data (see Figure~\ref{fig:dendro_leaves}). Solid line is CO, dashed line is \tco\ and dotted line is \cdo.  }
    \label{fig:leaf_intensities}
\end{figure}

We extract the emission of the CO isotopologues within each \cdo\ leaf (see Table~\ref{t:leaves}) and plot the summed line profiles in Figure~\ref{fig:leaf_intensities}. 
In general, the most optically thick line, CO (shown with a solid line), has the broadest line profile. 
\tco\ (dashed line) and 
\cdo\ (dotted line) are each narrower, with some \cdo\ profiles showing evidence of multiple velocity peaks. 
The three leaves associated with jet-driving sources are 
L1 (HH~1066), L7 (HH~902), and L10 (HH~901). 
L1 shows clear evidence of two distinct velocity components within the beam separated by $\sim -2.65$~\kms. 
None of the three show evidence of red- and blue-shifted emission in the linewings that indicates an associated molecular outflow. 
These are likely washed out in the larger ACA beam as molecular outflows were detected in both HH~901 and HH~902 by \citet{cortes-rangel2020}.

We use the median \cdo\ line profile of each leaf to estimate its virial mass. 
We compute the virial mass using the following equation:
\begin{equation}
    M_{vir} = \frac{3(5-2n)}{8(3-n)\ln(2)} \frac{(\Delta v^2)R}{G}
\end{equation}
where 
$n$ is the exponent of the density profile ($\rho \propto r^{-n}$; we assume $n=2$),  
$\Delta v$ is the \cdo\ linewidth (FWHM), 
$R$ is the mean leaf radius, and
$G$ is the gravitational constant.  
Further corrections to account for the non-spherical morphologies will change these values by $<10$\% \citep{bertoldi1992}. 
Virial masses, along with the linewidths, and 1D velocity dispersions for each leaf are reported in Table~\ref{t:leaves}. 
We compare this to the molecular mass of each clump, computed as in Section~\ref{ss:alma_masses} using $\left[\mathrm{H_2}/\mathrm{C^{18}O} \right] = \left[\mathrm{H_2}/\mathrm{CO} \right] \times \left[\mathrm{^{12}CO}/\mathrm{C^{18}O} \right] = 6.16 \times 10^{6}$ using data from \citet{wilson1994} and \citet{pineda2010}. 
The molecular mass of all clumps is higher than their virial mass, suggesting that they unstable to collapse, as noted in Table~\ref{t:leaves}.

\subsubsection{Continuum sources}\label{ss:continuum}
\begin{figure*}
	\includegraphics[width=0.95\columnwidth]{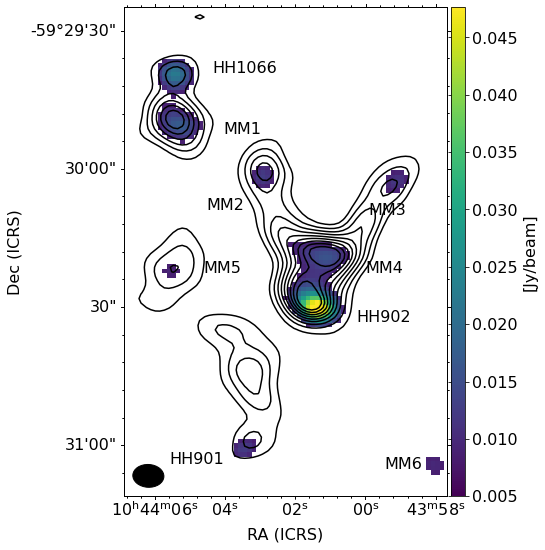}
    \includegraphics[width=0.95\columnwidth]{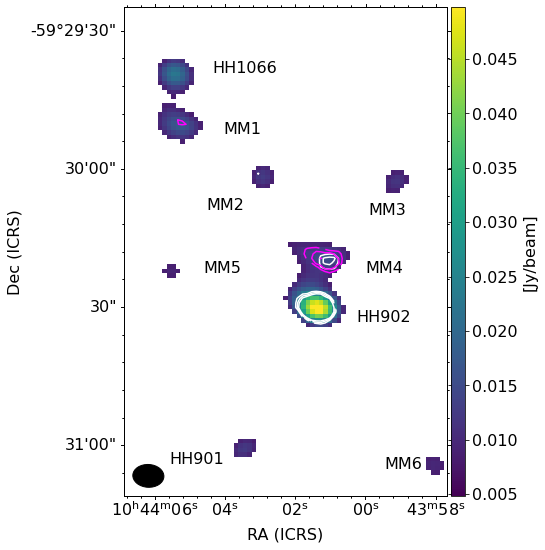}
    \caption{
    \textbf{Left:} Colorscale of the aggregate continuum with \cdo\ moment~0 contours from 10-100$\sigma$ in steps of 10$\sigma$.
    \textbf{Right:} Contours of the rarer molecules in our sample plotted on a colorscale of the continuum intensity. White contours show the DCN J=3-2 from 3-5~$\sigma$ and the magenta contours show$^{13}$CS J=5-4 from 1--3~$\sigma$. }
    \label{fig:cont_contours}
\end{figure*}

Figure~\ref{fig:cont_contours} shows 
continuum emission detected with significance $\gtrsim 5\sigma$.
This reveals 9 point sources. 
We detect continuum from clumps near the origin of all three of the famous HH jets in the \mm\ -- HH~901~MM, HH~902~MM, and HH~1066~MM. 
Five other continuum peaks fall within the \mm\ (MM1-MM5); three of these reside in the HH~902 pillar (MM2-4; see Figure~\ref{fig:cont_contours}). 
The final point source, MM6, lies outside the \mm\ complex so we do not discuss it further.

All mm continuum sources are associated with a \cdo\ leaf, 
although not all \cdo\ leaves have a continuum detection. 
Continuum and \cdo\ emission are well-aligned spatially in all leaves but L7 at the head of the HH~902 pillar. 
Using a 2D Gaussian fit, we determine an offset of $2.7\arcsec\pm 0.1\arcsec$ (0.03~pc) between the \cdo\ and continuum peaks. 
The continuum peak is offset toward the western side of the pillar, in the same direction as the HH~902~YSO seen at higher resolution by \citet{cortes-rangel2020}. 
A second continuum source detected to the northeast of the HH~902~YSO, HH~902~B, is not resolved with our larger beam.

Three additional continuum sources reside further north in the HH~902 pillar. 
A distinct peak in the continuum and \cdo\ emission traces MM4 immediately to the north of the HH~902~MM. 
A bridge of continuum emission connects the two sources. 
Two additional continuum sources lie further north in the wishbone-shaped HH~902 pillar. 
MM2 and MM3 coincide with \cdo\ peaks at the east and west tips of the pillar.

To the east of the HH~902 pillar, there are three continuum sources in the HH~1066 pillar. 
At the top (northernmost point) of the HH~1066 pillar is HH~1066~MM. 
Due south lies MM1. 
Like its neighbor, continuum and \cdo\ emission from MM1 appear to peak behind a cloud edge traced by bright H$\alpha$ emission. 
Further south, MM5 coincides with a \cdo\ peak at the head of the HH~1066 pillar, adjacent to the continuum emission from the HH~902~MM and MM4.

Finally, at the tip of the \mm\ complex, the HH~901 pillar has one continuum source detected at the head of the pillar. 
The continuum source overlaps with a local peak in the \cdo\ emission. 
However, the brightest \cdo\ emission in the HH~901 pillar is seen at its center where there is no continuum detection.

We compute masses of the continuum sources as: 
\begin{equation}
    M_d = S_{\nu} d^2 / B_{\nu}(T) \kappa_{\nu}. 
\end{equation}
We use a dust absorption coefficient 
$\kappa_{\nu}$ = 0.8~cm$^2$g$^{-1}$ \citep{oss94} 
appropriate for 1.3~mm observations, 
and a temperature of T=30~K. 
We multiply the dust mass by a gas-to-dust ratio of 100 and report the total mass in Table~\ref{t:fluxes_continuum}. 
We have assumed a single temperature for all sources regardless of differences in their evolutionary stages (i.e., the presence of jets). 
Assuming a higher or lower temperature (15~K or 45~K) changes the estimated mass by a factor of $\sim 2$. 
We refrain from a more detailed analysis given the low resolution of our data and evidence that multiple sources are unresolved in the beam \citep[note that][detect at least two point sources in the same region as our HH~902~MM]{cortes-rangel2020}. 
The flux and mass estimates of all continuum point sources are reported in Table~\ref{t:fluxes_continuum}. 

%%----------------------------------------------------------------
\begin{table}
\caption{Properties of the continuum sources.\label{t:fluxes_continuum}}
\centering
\begin{footnotesize}
\begin{tabular}{lllrr}
\hline\hline
Source & R.A. & decl.&  225.9~GHz$^{\dagger}$ & Mass \\
        & (J2000) & (J2000) & [mJy] & [\Msun] \\
        \hline
HH~1066~MM & 10:44:05.421 & $-$59:29:39.81 & 18.1 & 1.5 \\ 
HH~902~MM  & 10:44:01.415 & $-$59:30:29.84 & 65.9 & 5.3 \\ 
HH~901~MM  & 10:44:03.463 & $-$59:31:00.89 & 6.43 & 0.52 \\ 
MM1        & 10:44:05.327 & $-$59:29:50.65 & 13.3 & 1.1 \\ 
MM2        & 10:44:02.939 & $-$59:30:01.92 & 5.35 & 0.43 \\ 
MM3        & 10:43:59.146 & $-$59:30:02.95 & 6.25 & 0.51 \\ 
MM4        & 10:44:01.255 & $-$59:30:19.23 & 14.3 & 1.2 \\ 
MM5        & 10:44:05.503 & $-$59:30:22.43 & 4.76 & 0.39 \\ 
MM6        & 10:43:58.069 & $-$59:31:03.97 & 6.97 & 0.57 \\ 
\hline
\multicolumn{5}{l}{$^{\dagger}$Aggregate continuum from the Band~6 observations.}\\
\end{tabular}
\end{footnotesize}
\end{table}
%%----------------------------------------------------------------

\subsubsection{Candidate YSOs detected in other surveys}\label{ss:other_ysos}

Previous surveys at wavelengths $<$1.3~mm have also reported candidate YSOs in and around the \mm. 
Three candidate YSOs from the Pan-Carina YSO Catalog \citep[PCYC;][]{povich2011} fall within the area of our ALMA map; these are shown on Figure~\ref{fig:dendro_leaves} as black diamonds. 
All three candidate YSOs have an ambiguous evolutionary classification.  
PCYC~429 was identified as the HH~1066 driving source by \citet{reiter2016} and coincides with the continuum source HH~1066~MM and the \cdo\ emission of L1. 
Immediately below HH~1066~MM, a second candidate YSO, PCYC~427, lies near the northwest boundary of L2, outside the continuum emission of MM1.  
PCYC~427 coincides with a point source visible in H$\alpha$ images suggesting that this source lies in front of the cloud. 
Further west, a third candidate YSO, PCYC~399, is also visible in H$\alpha$ images. 
This source falls well outside the lowest contour of CO emission from the HH~902 pillar and has no associated continuum emission, suggesting that it is also lies outside the cloud.

\citet{ohl12} searched for YSOs driving the prominent jets in the \mm\ and other dust pillars in Carina. 
They note a point-like source near HH~902 that is not detected at longer ($>8$\micron) wavelengths; this is shown as a cyan square in Figure~\ref{fig:dendro_leaves}. 
This source falls within HH~902~MM but is offset to the left, closer to the peak of the \cdo\ emission. 
Coordinates reported by \citet{ohl12} place the source $\sim4\arcsec$ away from the HH~902~YSO seen with ALMA by \citet{cortes-rangel2020}. 
Together this suggests at least 3 YSOs unresolved in the HH~902~MM. 
Finally, no \emph{Herschel} point sources were identified in the Mystic Mountains in the unbiased search by \citet{gaczkowski2013}.

\subsection{Rarer species}\label{ss:rare}
\begin{figure*}
  \includegraphics[width=\columnwidth]{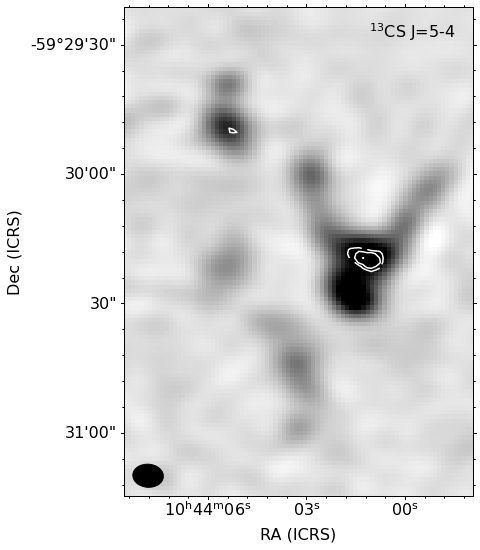}
  \includegraphics[width=\columnwidth]{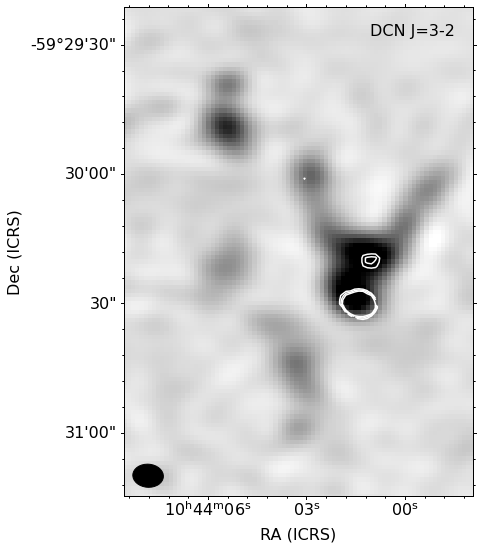}
    \caption{Moment~0 map of the \cdo\ (grayscale) with contours of the $^{13}$CS J=5-4 in steps of 1~$\sigma$ from 1--3~$\sigma$ (\textit{left}) and 
    DCN J=3-2 emission in steps of 1~$\sigma$ from 3--5~$\sigma$ (\textit{right}). }
    \label{fig:other_lines}
\end{figure*}

Our observations included two rarer molecules, $^{13}$CS J=5-4 and DCN J=3-2. 
$^{13}$CS is a high-density tracer, typically seen in molecular envelopes and cavity walls \citep{tychoniec2021}. 
DCN is most often observed in cold gas, where CO is frozen out of the enabling deuterium chemistry \citep{caselli2012}. 
Contour plots showing the emission peaks of both lines are presented in Figure~\ref{fig:other_lines} and Table~\ref{t:rarer_props}.
The strongest $^{13}$CS J=5-4 emission is associated with MM4. 
A second, weaker feature coincides with MM1.

DCN J=3-2 was serendipitously detected in a spectral window centered on SiO J=5-4;  
we do not detect SiO J=5-4 in any part of the mosaic. 
The brightest DCN peak is at the head of the HH~902 pillar, coincident with the HH~902~MM continuum source. 
Like the continuum emission, DCN is offset to the west of the \cdo\ emission (see Figure~\ref{fig:cont_contours}).

\begin{figure}
	\includegraphics[width=\columnwidth]{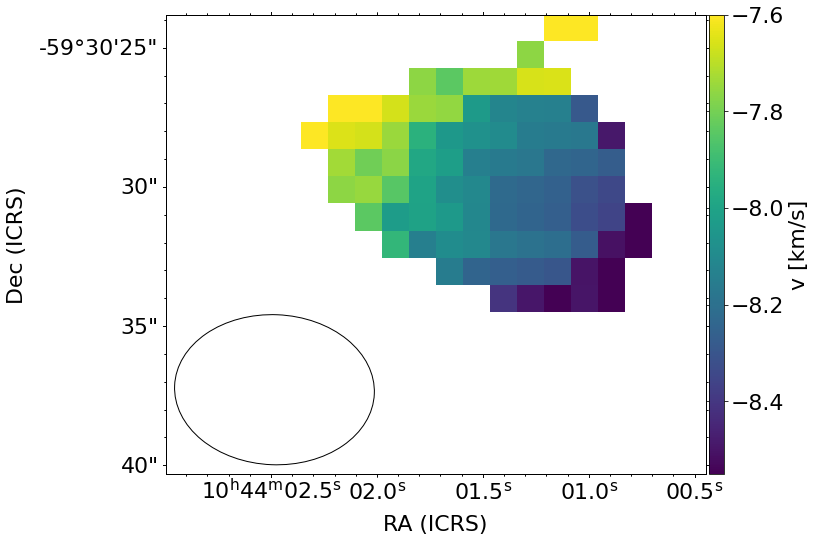}
	\includegraphics[width=\columnwidth]{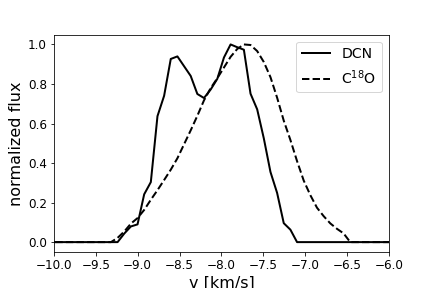}
    \caption{
    \textbf{Top:} Moment~1 map of the DCN J=3-2 emission at the tip of the HH~902 pillar (see Figure~\ref{fig:other_lines}) that shows the velocity gradient across the marginally resolved source (the beam is shown in the lower left). 
    \textbf{Bottom:} Spectrum of the summed DCN J=3-2 emission compared to \cdo\ spectrum from the same region. 
    }
    \label{fig:dcn_vel}
\end{figure}

DCN emission coincident with the HH~902~MM presents a double-peaked velocity profile, shown in  Figure~\ref{fig:dcn_vel}. 
The red peak is within $0.2$~\kms\ of the velocity peak of the \cdo\ emission from the same region. 
The second peak is blueshifted by $\sim$1~\kms. 
This velocity separation is not consistent with the frequency of any of the hyperfine components of DCN \citep[using data from the Cologne Database for Molecular Spectroscopy; CDMS;][]{mueller2001,mueller2005}. 
The flux of both peaks is remarkably similar, contrary to expectation for optically thin hyperfine emission.  
Finally, the velocity gradient is marginally resolved across the HH~902~MM (see Figure~\ref{fig:dcn_vel}). 
\citet{cortes-rangel2020} observed a similar velocity gradient in N$_2$D$^+$ J=3-2 (note that their observations have $\sim20\times$ better spatial resolution). 
Both deuterated molecules show a velocity gradient that is oriented \textit{opposite} to the velocity structure of the HH~902 molecular outflow -- that is, redshifted DCN (N$_2$D$^+$) is seen on the same side of the YSO as the blueshifted limb of the outflow.

A second, fainter DCN peak coincides with MM4. 
The emission peaks of all three molecules -- DCN, $^{13}$CS, and \cdo\ -- coincide with continuum emission from MM4. 
Weak DCN emission extends from the position of MM2 along the edge of the pillar (see Figure~\ref{fig:mom0_maps}).

%

%%======================================================================
\begin{table}
\caption{Summary of the rarer molecules. Columns are the species/transition, peak and median intensities, and median optically-thin column density. \label{t:rarer_props}}
\centering
\begin{footnotesize}
\begin{tabular}{rllll}
\hline\hline
source & $I_{\mathrm{peak}}$ &
$I_{\mathrm{median}}$ & 
 log(N)$_{\mathrm{peak}}$ & 
 log(N)$_{\mathrm{median}}$\\ 
        & [K~km~s$^{-1}$] & [K~km~s$^{-1}$] \\
\hline
\multicolumn{5}{c}{DCN J=3-2 }\\
\hline
MM~HH902 & 0.812 & 0.310  & 12.5 & 12.1 \\
MM~4 & 0.298 & 0.116  & 12.1 & 11.7 \\
\hline
\multicolumn{5}{c}{$^{13}$CS J=5-4 }\\
\hline
MM~4 & 0.355 & 0.136 & 12.3 & 11.9 \\
MM~1 & 0.121 & 0.076 & 11.8 & 11.6 \\
\hline
\end{tabular}
\end{footnotesize}
\end{table}
%
%%==========================================================================

%---------------------------------------------
\subsection{Velocity structure}\label{ss:fedback_vel}
%---------------------------------------------
%
\begin{figure*}
    \includegraphics[scale=0.275]{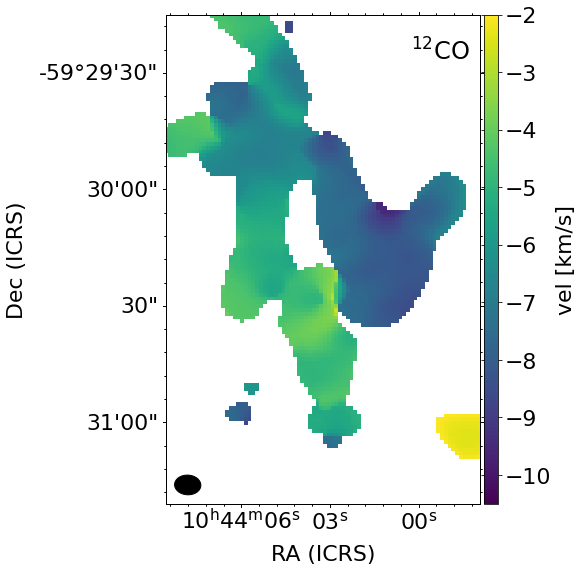}
	\includegraphics[scale=0.275]{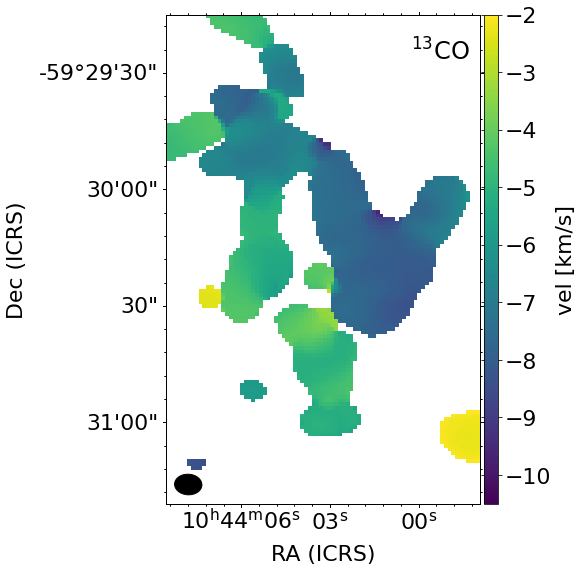}
	\includegraphics[scale=0.275]{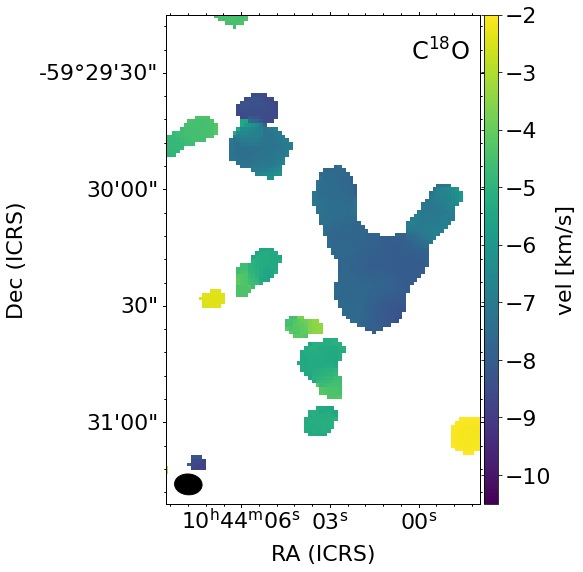}
	\caption{Intensity-weighted velocity (moment~1) maps of the CO isotopologues. }
    \label{fig:velocities}
\end{figure*}

Intensity-weighted velocity (moment~1) maps of the CO isotopologues are shown in Fig.~\ref{fig:velocities}. 
The most prominent feature is the velocity difference between the pillars of the \mm. 
Some velocity differences are also apparent along the north/south axis of the HH~1066 pillar and the wishbone-like extensions of the HH~902 pillar. 
\cdo\ emission suggests that there are multiple clumps in the HH~1066 pillar with slightly different velocities. Higher resolution observations are required to determine if the inter-clump gas is at a markedly different velocity, as seen in Pillar~6 of \citet{klaassen2020}.

%---------------------------------------------
\subsection{Non-thermal motions}\label{ss:non_therm}
%---------------------------------------------

\begin{figure}
    \centering
    \includegraphics[width=\columnwidth]{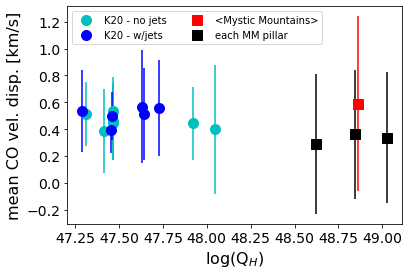}
    \caption{Mean CO velocity dispersion compared to the estimated incident ionizing photon luminosity for the three pillars in the \mm\ (black squares)
    and the average of the \mm\ (red square). 
    These are compared to other pillars in Carina observed with the same resolution and spectral setup from \citet{klaassen2020} with (blue) and without (cyan) one or more prominent jets seen at visual wavelengths. All three pillars in the \mm\ complex have at least one jet. }
    \label{fig:vel_disp_comp}
\end{figure}

To test whether photoionization contributes to the non-thermal motions in the pillars, we compare the average velocity dispersion to the incident ionizing photon flux. 
To probe a larger range of fluxes, we compare the \mm\ to the dust pillars in Carina from \citet{klaassen2020} because those observations use the same angular resolution and spectral setup as this work. 
As in \citet{klaassen2020}, we compute the average velocity dispersion from the moment~2 map. 
We consider the \mm\ as a whole and each of the three pillars within it separately. 
To calculate the incident ionizing photon flux, we assume that the three main star clusters, Tr14, Tr15, and Tr16, dominate the external irradiation for all pillars in Carina. 
This provides a lower bound as we neglect the large number of O- and B-type stars located outside these clusters \citep[see, e.g.,][]{berlanas2023} and the effects of extinction. 
We use the ionizing photon luminosities from \citet{smith2006_energy} and compute the local flux using the projected distance between the pillars and the clusters\footnote{The nearest cluster in projection does not always dominate the ionization. For example, Pillar~20 lies alongside Tr15 but points toward Tr16, suggesting that it has had a stronger influence. }. 
The mean pillar velocity dispersion as a function of incident ionizing photon flux is shown in Figure~\ref{fig:vel_disp_comp} (color-coded by whether the pillars contain a jet seen at visual wavelengths, see discussion in Section~\ref{ss:carina_comp}).

Within the (large) uncertainties, the mean velocity dispersions are similar across nearly two orders of magnitude in incident ionizing radiation. 
This is counter to expectation if ionizing radiation drives turbulence in the gas. 
However, moment~2 maps do not isolate non-thermal motions from other factors that contribute to the linewidth. 
Bulk motions from large-scale processes like infall and rotation as well as the influence of outflows from
protostars embedded in the pillars will all contribute to this single value. 
In addition, higher temperatures will increase the thermal contribution to the linewidth. 
However, to first order we would expect pillars subject to higher ionizing photon fluxes would also have warmer temperatures 
\citep[see, e.g., the dust temperature maps from][]{roccatagliata2013,rebolledo2016}.

\begin{figure*}
	\includegraphics[scale=0.275]{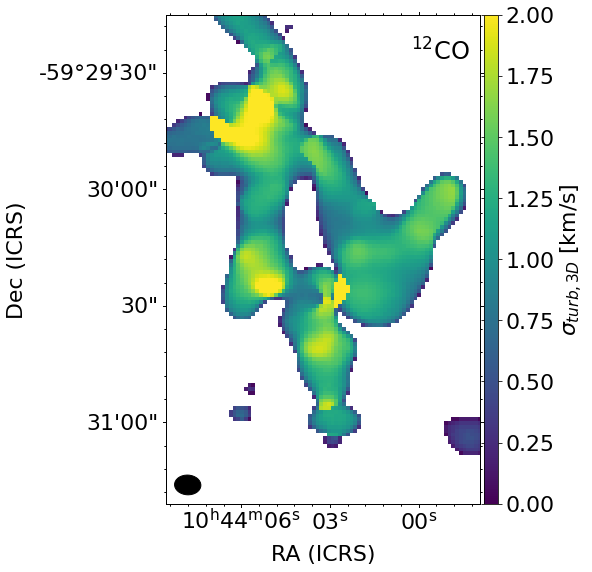}
	\includegraphics[scale=0.275]{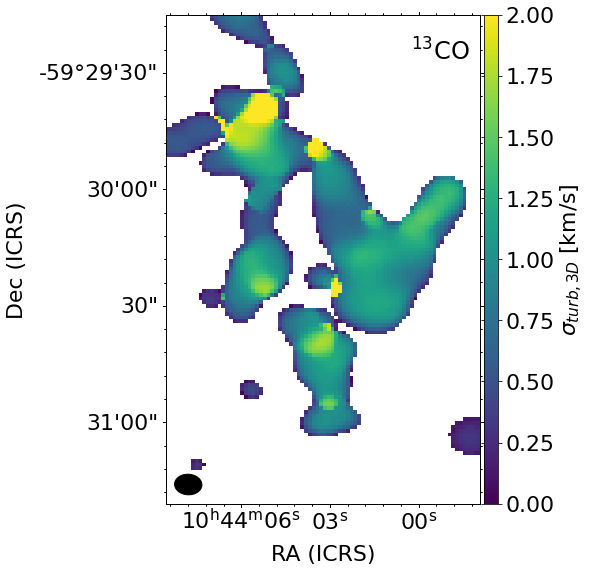}
	\includegraphics[scale=0.275]{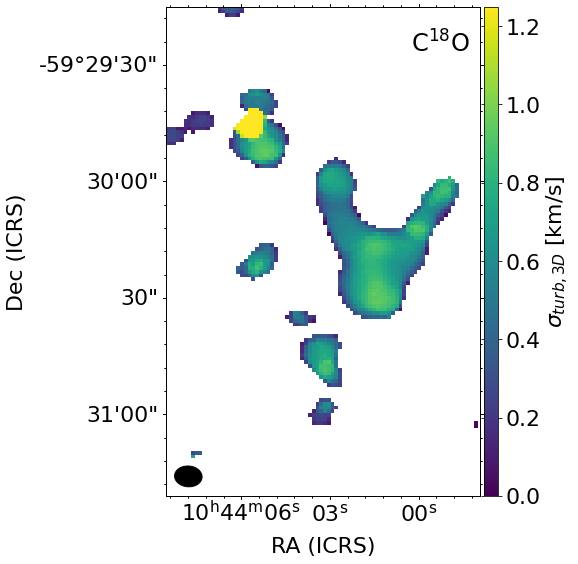}
	\caption{The 3D turbulent velocity dispersion computed assuming a single gas temperature of $T=30$~K. 
	}
    \label{fig:sigmas}
\end{figure*}

For the \mm\ we separate thermal and turbulent motions as follows. 
We use the linewidth measured as the full width at half maximum (FWHM; $\Delta v$) to calculate the 1D velocity dispersion, 
$\sigma = \Delta v/2 \sqrt{2 \ln{2}}$. 
To compute the thermal contribution, we use 
$\sigma_{\mathrm{therm;1D}} = \sqrt{2 k_B T / m_{\mathrm{iso}}}$
where $m_{\mathrm{iso}}$ is the mass of the CO isotopologue used and T=30~K is the temperature of the cold molecular gas. 
Subtracting the thermal component from the total velocity dispersion yields the 1D turbulent velocity
dispersion, $\sigma_{\mathrm{turb;1D}} = \sqrt{\sigma_{\mathrm{tot}}^2 - \sigma_{\mathrm{therm}}^2}$. 
Finally, we convert the 1D estimate to 3D as $\sigma_{\mathrm{turb}} = \sqrt{3} \sigma_{1D}$. 
The 3D turbulent velocity dispersion computed for each CO isotopologue is shown in Figure~\ref{fig:sigmas}. 

\begin{figure*}
	\includegraphics[scale=0.405]{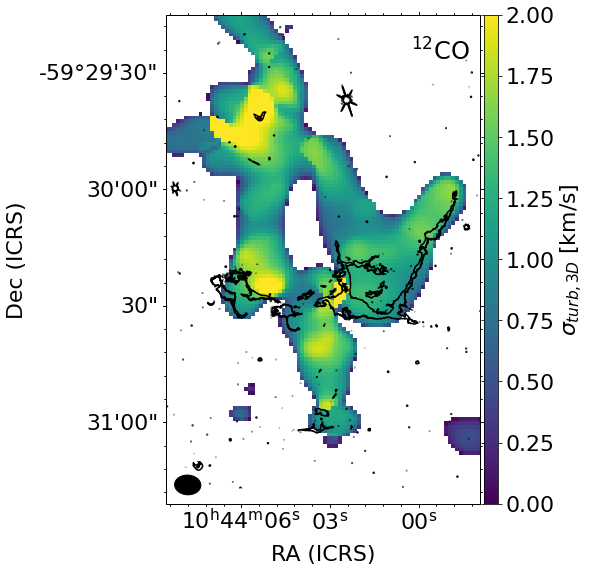}
	\includegraphics[scale=0.405]{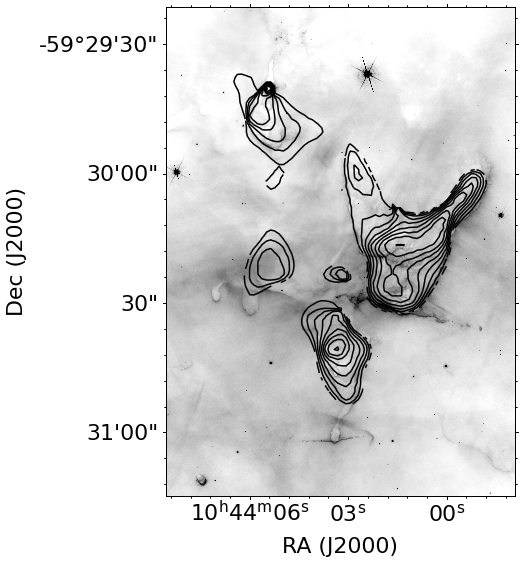}
	\includegraphics[scale=0.405, trim=0mm 0mm 
	10mm 0mm]{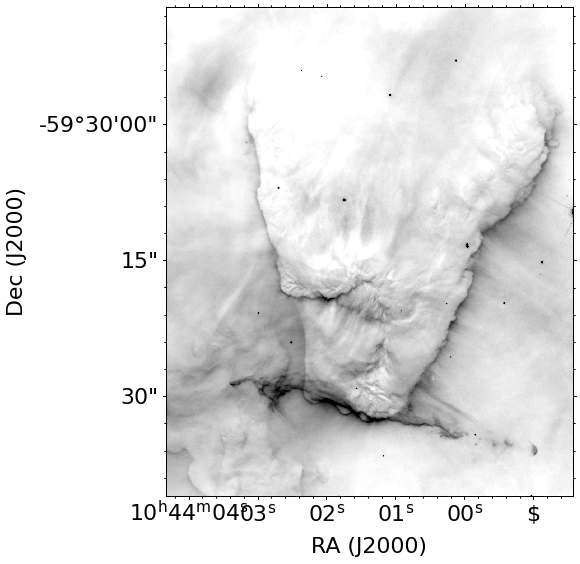}
	\includegraphics[scale=0.405, trim=10mm 0mm 0mm 0mm]{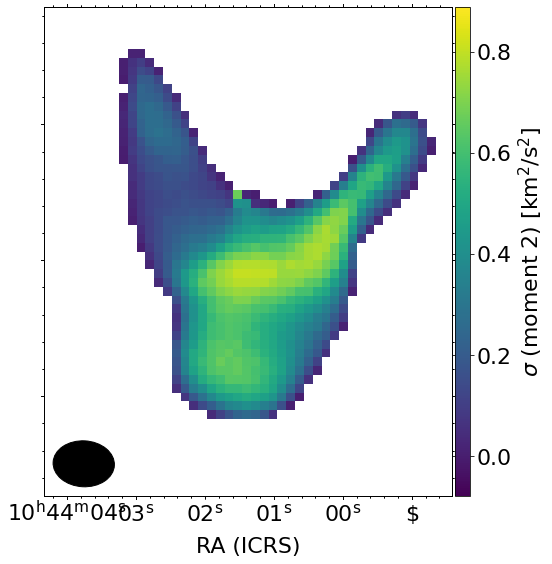}
	\caption{
	\textbf{Top Left:} Map of the CO turbulent velocity dispersion with H$\alpha$ contours overplotted in black. 
    \textbf{Top Right:} Contours from the CO moment 2 map plotted on a grayscale of the H$\alpha$ image from \emph{HST}.  Many peaks in the velocity dispersion correspond to regions where the pillars appear to overlap. The one exception is HH~902 which we zoom in on in the bottom two panels. 
	\textbf{Bottom:} A zoomed-in view of the HH~902 pillar showing the \emph{HST} H$\alpha$ image alongside the CO moment 2 map. Both the H$\alpha$ intensity and the CO velocity dispersion are higher along the western side of the HH~902 pillar. 	}
    \label{fig:hh902_sigmas}
\end{figure*}

Velocity dispersions are higher in CO than in \tco\ and \cdo. 
The highest values for the turbulent velocity dispersion are observed where the pillars overlap, almost certainly reflecting the influence of multiple velocity components along the line of sight. 
Other regions with high values of the velocity dispersion fall behind ionization fronts traced by H$\alpha$ (see Figure~\ref{fig:hh902_sigmas}).

The HH~902 pillar is well resolved with our $\sim$6\arcsec\ beam and overlaps only minimally with the other pillars in the \mm. 
Qualitatively, the velocity dispersion is higher where the H$\alpha$ emission is higher. 
Velocity dispersions are larger along the western edge of the pillar and modest peaks appear behind the two H$\alpha$ ridges at the head of the pillar. 
Peaks in the velocity dispersion do not coincide with the location of the continuum sources.

When estimating $\sigma_{\mathrm{therm;3D}}$, we assumed a single temperature of T=30~K for the entire \mm. 
Gas temperatures are likely higher in the most heavily irradiated parts of the \mm. 
Assuming a higher temperature, T=50~K, yields a larger thermal contribution $\sigma_{\mathrm{therm;1D}} = 0.59$~\kms\ which is roughly equivalent to the average velocity dispersion of the \mm\ (as estimated from the moment~2 map). 

In addition to temperature uncertainties, the lack of correlation between velocity dispersion and incident ionizing photon flux may result from one of the following possibilities: 
(1) turbulence is enhanced only near irradiated cloud surfaces in layers too thin to be resolved with our $\sim$6\arcsec\ beam; 
(2) different pillars have a different level of embedded star formation that contributes unequally to the observed linewidth (i.e. via outflows) but is unresolved in our beam (and star formation activity itself does not vary as a function of incident ionizing flux); or 
(3) that ionizing radiation does not drive turbulence.

However, the velocity dispersion may not be best discriminant of the impact of ionizing radiation on the gas. 
External radiation may drive non-thermal motions through compressive shocks but shock energy will quickly dissipate and the velocity signature may not be long-lived. 
Compressive motions will also reshape the density of the gas, leaving behind local regions of higher density. 
We consider this signature of compressive turbulence in the next section.

\subsection{Isolating turbulent motions}\label{ss:turb_motions}

\citet{menon2021} used the data of \citet{klaassen2020} to measure the turbulence in a few well-resolved pillars. 
They reconstruct the dominant turbulence driving mode from the column density and intensity-weighted velocity maps \citep{federrath2010,federrath2016}. 
In this section, we repeat this analysis for the \mm\ complex for a highly irradiated comparison. \\

%------------------------------------------
\noindent \textbf{2D density structure:}
%------------------------------------------
To compute the density  probability distribution
function (PDF), we use the column density of the optically thin \tco\ as a proxy for the H$_2$ along the line of sight, $N$, substituting the \cdo\ in places where the \tco\ is optically thick.  
We compute $\sigma_{\eta}$, the 
dispersion of the natural logarithm of the column density scaled by its mean ($\eta= \log(N/N_0)$) by fitting a \citet{hopkins2013} intermittency density PDF model to the volume-weighted PDF of $\eta$. 
This has the form 
\begin{equation}
    p_{\mathrm{HK}}(\eta)d\eta = I_1 ( \sqrt{2 \lambda \omega(\eta)} \exp{[-(\lambda + \omega(\eta))]} \sqrt{ \frac{\lambda}{\theta^2 \omega(\eta)}} d\eta
\end{equation} 
where
\begin{equation}
    \lambda = \frac{\sigma_{\eta}^2}{2\theta^2}
\end{equation}
and 
\begin{equation}
    \omega(\eta) = \frac{\lambda}{(1+\theta)} - \frac{\eta}{\theta} \,\,\, (\omega \geq 0)
\end{equation}
and 
$\theta$ is the intermittency parameter.  
The values of $\sigma_{\eta}$ and $\theta$ derived from the fit are used to compute 
$\sigma_{N/N_0}$, the  linear dispersion, as 
\begin{equation}
    \sigma_{N/N_0} = \sqrt{ \exp{ \left(\frac{\sigma_{\eta}^2}{(1 + 3\theta + 2\theta^2)} \right)} - 1}
\end{equation}
using an expression from \citet{hopkins2013}. \\

%------------------------------------------
\noindent \textbf{Conversion from 2D to 3D density structure:}
%------------------------------------------
We estimate the 3D density dispersion from the 2D column density dispersion using the method of \citet{brunt2010}. 
The 3D density power spectrum, P$_{3D}(k)$, of the variable $\rho/\rho_0 -1$ is reconstructed from the 2D column density power spectrum, P$_{2D}(k)$, of the variable $N/N_0 - 1$; $k$ is the wavenumber. 
This is converted to the 3D power spectrum as 
P$_{3D}(k) = 2 k \mathrm{P}_{2D}(k)$. 
The ratio $\mathcal{R}^{1/2}$ of the 2D and 3D dispersions is defined as 
\begin{equation}
    \mathcal{R}^{1/2} = \frac{\sigma_{N/N_0}}{\sigma_{\rho/\rho_0}} = \frac{\sum_k P_{2D}(k)}{\sum_k P_{3D}(k)}
\end{equation}
where we have mirrored the column density to provide a periodic dataset \citep{ossenkopf2008}. 
\\

%------------------------------------------
\noindent \textbf{Isolating turbulent motions:}
%------------------------------------------
We fit a plane to the intensity-weighted velocity map (the first moment map) to remove bulk motions and isolate turbulent motions in the pillars. 
For purely turbulent motion, we expect the line of sight motions to trace a Gaussian PDF. 
We fit a Gaussian to the line of sight velocity PDF to derive the 1D velocity dispersion, $\sigma_{v,1D}$. 
We convert this to the 3D velocity dispersion as 
$\sigma_{v,3D} = 3^{1/2} \sigma_{v,1D}$, 
implicitly assuming isotropy.

From this, we compute the Mach number, $\mathcal{M} = \sigma_{v,3D} / c_s$, 
which is the ratio of the 3D velocity dispersion and the sound speed, $c_s \sim 0.3$~\kms\ for our assumed temperature of 30~K. 
Finally, we compute $b$, the turbulence driving parameter as $b= \sigma_{\rho/\rho_0} / \mathcal{M}$. 
The value of $b$ is used to determine the type of turbulence: 
$b \sim 0.33$ is purely solenoidal;  
$b \sim 1.0$ is purely compressive; and  
$b \sim 0.4$ is a combination of both.

\begin{figure*}
    \centering
    \includegraphics[width=0.355\textwidth,trim=0mm -10mm 0mm 0mm]{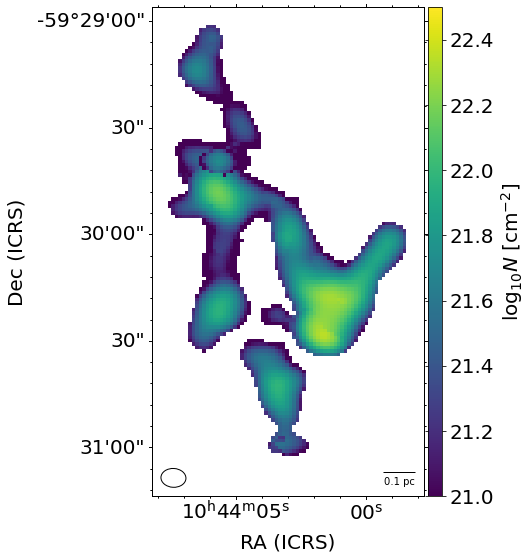}
    \includegraphics[width=0.605\textwidth]{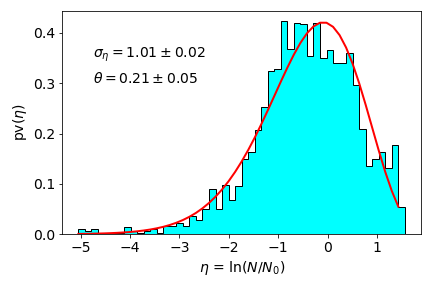}
    \includegraphics[width=0.355\textwidth,trim=0mm -10mm 0mm 0mm]{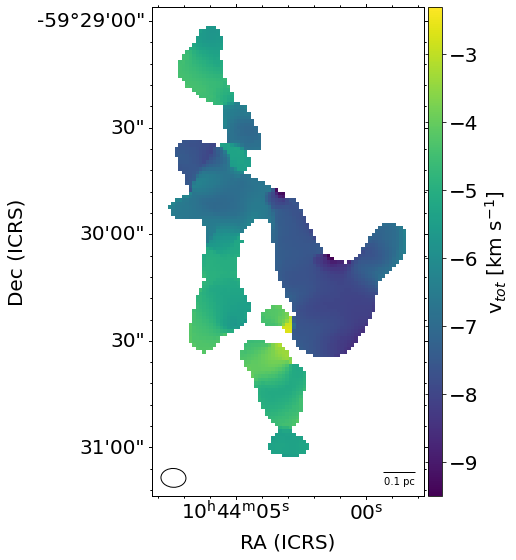}
    \includegraphics[width=0.605\textwidth]{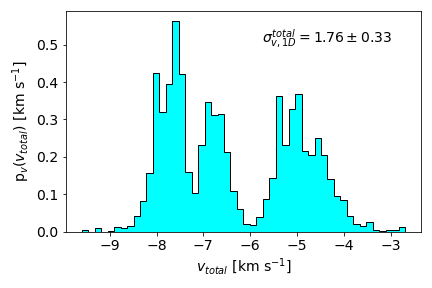}
    \includegraphics[width=0.355\textwidth,trim=0mm -10mm 0mm 0mm]{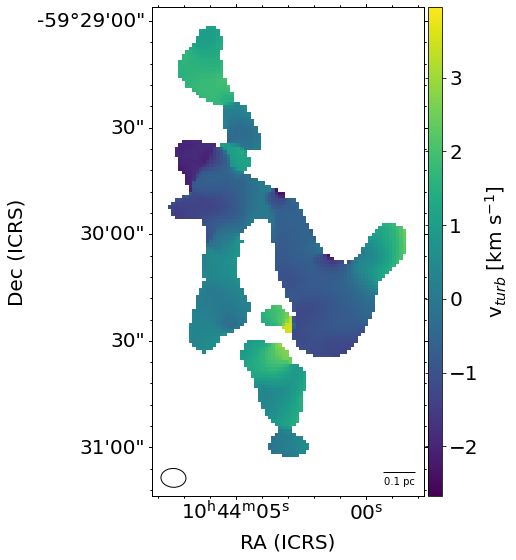}
    \includegraphics[width=0.605\textwidth]{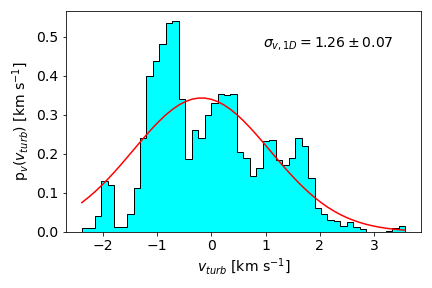}
    \caption{\textbf{Top:} Left panel shows the H$_2$ column density derived from the optically thin \tco\ and \cdo\ (see Section~\ref{ss:non_therm}). Right panel shows the fit to the column density PDF, as in \citet{menon2021}. 
    \textbf{Middle:} Image of the intensity-weighted velocity map prior to gradient subtraction (left) and histogram of its distribution (right). 
    \textbf{Bottom:} Velocity map (left) and histogram (right) after gradient subtraction. }
    \label{fig:turb_derv}
\end{figure*}

Values of each of the derived parameters and their 1$\sigma$ uncertainties 
are listed in Table~\ref{t:turb_params}.  
PDFs of the column density and velocities for the \mm\ are shown in Figure~\ref{fig:turb_derv}. 
We focus our comparison on the \mm\ as a whole because the individual pillars are significantly smaller, with only a few beams covering their major and minor axes. 
Results of this analysis for each individual pillar are shown in Appendix~\ref{appendix:pillar_vturb} but we do not discuss them further here as the individual pillars are inadequately resolved for this analysis 
(similar to Pillar~44 in \citealt{menon2021}; see also \citealt{sharda2018}). 
\\

%------------------------------------------
\noindent \textbf{Comparing the \mm\ to other pillars in Carina:}
%------------------------------------------
Within the \mm, it is clear that there are multiple velocity components; 
these roughly correspond to the systemic velocity of each pillar (see Figure~\ref{fig:turb_derv} and Table~\ref{t:molecular_props}). 
Subtracting a linear function (a plane) reduces the peakiness of the velocity distribution, but the velocity dispersion remains large even after gradient subtraction, nearly a factor of two higher than the values found by \citet{menon2021} for other pillars in Carina. 
As a result, the Mach number ($\mathcal{M}=\sigma_{v,3D}/c_s$) is also a factor of 2 higher. 
The resulting turbulence driving parameter b is within the range of compressively-dominated turbulence ($0.4-1.0$) but somewhat lower than the values
($0.8-1.7$) found by \citet{menon2021}.

%%----------------------------------------------------------------
\begin{table}
\caption{Turbulence parameters for the \mm\ as a whole, as in \citet{menon2021}. \label{t:turb_params}}
\centering
\begin{footnotesize}
\begin{tabular}{cc}
\hline
Pillar & MM \\ 
\hline
A [pc$^2$] & 0.32$\pm 0.03$ \\
$N_0$ [10$^{21}$ cm$^{-2}$] & 3.01$^{+5.3}_{-1.9}$ \\%[1.5mm]
$M$ [\Msun] & 33.7$\pm 5.7$ \\
n [$10^3$cm$^{-3}$] & 2.33$\pm 1.3$ \\
$\sigma^{M2}_{v,3D}$ [\kms] & 0.34$\pm0.30$ \\
$\alpha_{vir}$ & 26.7$\pm 15$ \\
t$_{\mathrm{ff}}$/t$_{\mathrm{turb}}$ & 2.22$\pm 0.62$ \\
$\sigma_{\eta}$ & 1.01$\pm 0.02$ \\
$\sigma_{N/N_0}$ & 0.56$\pm 0.02$ \\
$\mathcal{R}^{1/2}$ & 0.14$\pm 0.01$ \\
$\sigma_{\rho/\rho_0}$ & 4.02$\pm 0.37$ \\
$\sigma^{\mathrm{total}}_{v,1D}$ [\kms] & 1.76$\pm 0.33$ \\
$\sigma_{v,1D}$ [\kms] & 1.26$\pm 0.07$ \\
$\sigma_{v,3D}$ [\kms] & 2.19$\pm 0.13$ \\
$\mathcal{M}$ & 7.29$\pm 0.43$ \\
$b$ & 0.55$\pm 0.05$ \\
\hline 
\end{tabular}
\end{footnotesize}
\end{table}
%%----------------------------------------------------------------

%%--------------------------------------------------
% full MM 
%%--------------------------------------------------

%=======================================================================
\section{Discussion}
%=======================================================================
%

Recent numerical work suggests that ionizing radiation may drive turbulence in cold molecular gas via the rocket effect \citep[e.g.,][]{gritschneder2009,boneberg2015,dale2017,menon2020}. 
Photoevaporation drives a flow of free electrons from ionized cloud surfaces. 
Momentum is conserved, so this flow also drives a shock into the cloud, 
compressing the gas and perhaps injecting turbulence.

In this picture, highly irradiated pillars like the \mm\ should show a high level of turbulent velocity dispersion in the gas. 
The \mm\ cloud complex lies in the heart of the Carina Nebula where copious ionizing photons from Tr14 \citep[$Q_H \sim 10^{50}$~s$^{-1}$, see][]{smith2006_energy} 
illuminate and sculpt its mountainous morphology. 
The incident ionizing photon flux illuminating the \mm\ is an order of magnitude higher than that illuminating other pillars in Carina with similar observations \citep[see Figure~\ref{fig:vel_disp_comp} and][]{klaassen2020}.

A clear correlation between the incident ionizing radiation and electron density in ionization fronts has been observed in dust pillars in Carina \citep{mcleod2016}. 
We do not find a similarly straight-forward relationship between the cold molecular gas and incident radiation. 
Despite the higher ionizing flux illuminating the \mm, the average velocity dispersion (derived from the moment~2 map) is comparable to other pillars in Carina (see Figure~\ref{fig:vel_disp_comp}). 
At $b=0.55$, the turbulence driving parameter is consistent with compressive turbulence but is below the range found by \citet{menon2021}.

For all of the pillars in Carina that have been observed with the ACA alone, it is unclear what role unresolved star formation activity plays in the observed kinematics. 
In the \mm, we detect 
8 continuum sources (see Section~\ref{ss:continuum}) and 
10 \cdo\ clumps (see Section~\ref{ss:clumps_cores}). 
The pillars in the \citet{klaassen2020} also display a wide range of star-formation activity
from the relatively quiescent Pillar~22 which has only two \cdo\ cores to the actively star-forming Pillar~6 with evidence for $>$4 separate YSOs. 
Protostars and their (unresolved) outflows can contribute to internal velocity dispersion \citep[e.g.,][]{larson2015} and may provide a local source of turbulence. 
Quantifying this is difficult, however, as we do not detect the known outflows in the \mm\ or any other molecular outflows, likely due to the modest resolution of our data (at 2.3~kpc, our $\sim6$\arcsec\ beam corresponds to $\sim$0.07~pc). 
The physical properties of the famous jets in the \mm\ (see Figure~\ref{fig:hst_contours}) are well measured \citep{reiter2013,reiter2014,cortes-rangel2020} but these largely propagate outside the cloud.

Higher angular resolution observations are required to detect and remove the influence of embedded outflows as well as analyze the individual pillars of the \mm\ independently. 
Existing data covers a small region at the head of the HH~901 and HH~902 pillars \citep{cortes-rangel2020}; extending this to a larger portion of the \mm\ would be interesting to do in the future. 
For now, we compare the results of the \mm\ and the \citet{klaassen2020} pillars to higher resolution observations of other portions of the Carina region.

\subsection{Comparison to other regions in Carina }\label{ss:carina_comp}

A few other studies in the Carina region have quantified the impact of ionizing feedback on star-forming gas. 
\citet{rebolledo2020} compared two clouds in the Carina region that are subject to very different ionizing radiation fields: 
the `North Cloud' which is located near the center of Carina where it is heavily irradiated by Tr14 
and 
the `South Pillars' regions which is located in the outskirts of Carina where it is subject to much less intense radiation. 
The two regions are separated from the central clusters, Tr14 and Tr16, by $\sim$2.5~pc and $\sim$30~pc, respectively, and thus experience an order of magnitude difference in their radiative environment. 
With Band~3 observations targeting the dense gas tracers HCN and HCO$^+$ with a resolution of $2.8\arcsec\times\ 1.8\arcsec$,   
\citet{rebolledo2020} find evidence for more turbulence in the heavily irradiated North Cloud. 
The North Cloud also has fewer cores than the South Pillars region. 
However, the cores that have formed in the North Cloud are higher in mass than the more numerous cores found in the South Pillars cloud, consistent with turbulent fragmentation.

\citet{hartigan2022} recently published observations of the `Western Wall,' a region in the center of Carina that overlaps with the North Cloud from \citet{rebolledo2020}. 
These higher resolution observations (synthesized beamsize $\sim 1\arcsec$) in Band~6 target CO and its isotopologues.  
\citet{hartigan2022} conclude that the influence of feedback is modest with no signs of triggered star formation and no prominent dust pillars. 
Gas densities appear higher immediately behind the ionization front but cores appear starless, and there is no evidence for grain growth. 
A follow-up analysis of the \citet{hartigan2022} observations from \citet{downes2023} determined that turbulence is driven at large scales, but not necessarily by irradiation from nearby high-mass stars.

This conclusions stands in contrast to \citet{menon2021} who argue that predominantly compressive modes of turbulence may have triggered star formation in the pillars. 
However, \citet{downes2023} argue that their result is not in tension with \citet{menon2021}. 
Unlike the Western Wall, the pillars have been sculpted by radiation. 
Pillars may self-shadow, allowing compressive motions to lead to their development, altering their internal kinematics.

In the simulations of \citet{dale2012}, prominent pillars form in clouds with more diffuse gas and those with smoother density fields. 
Above a certain density ($\gtrsim$100~cm$^{-3}$), ionization is dynamically ineffective, especially in more turbulent regions. 
On the smaller scales of the pointed observations presented in this paper and those in \citet{klaassen2020}, \citet{rebolledo2020}, and \citet{hartigan2022}, differences in the local initial conditions may be responsible for the morphological differences observed at present. 
In this case, ionizing radiation carved more diffuse gas into dust pillars, perhaps triggering star formation \citep{brooks2002,rathborne2004,smith2010_spitzer,ohl12}. 
Meanwhile, the higher density Western Wall / North Cloud began as and remained a higher density region \citep[see, e.g., Figure~1 in][]{rebolledo2020}, less vulnerable to compression. 
Local density variations may also help explain why there is a dust pillar $\sim$1\arcmin\ ($\sim$0.7~pc) to the northeast of the Western Wall \citep[see][]{smith2010}.

This picture is in line with the simulations of \citet{tremblin2012_sc,tremblin2012_sct} who find that pillar formation depends strongly on shock curvature. 
Pre-existing overdensities help curve the shock driven by the ionization front. 
Pillars form when the curved shock collapses in on itself. 
Pillars formed in this way will naturally have a high-density core at their tips in much the same way that desert buttes have high-density caprock at their apices.

While differences in the initial density may help explain the absence of pillars in the Western Wall, the low level of star formation activity remains a challenge. 
High initial densities in the Western Wall suggest that star formation would happen sooner than in the feedback-carved pillars. 
But this is not what is observed. 
\citet{hartigan2022} find that the cores in the Western Wall are starless. 
\citet{rebolledo2020} argue that the higher level of turbulence may have made the Western Wall more resilient to fragmentation as there are fewer but higher mass cores compared to their more quiescent South Pillars region. 
In contrast, several pillars show evidence for star formation in the form of their 
prominent jets.

Turbulence rapidly decays and must be constantly resupplied to maintain observed levels. 
\citet{rebolledo2020} attribute the higher turbulence in the North Cloud / Western Wall to the impact of external irradiation. 
\citet{downes2023} also find evidence that turbulence is driven on large scales, but they do not attribute this to ionization. 
Resolving this tension requires a more homogeneous dataset that covers the large (pc) scales where turbulence is driven while resolving the 0.02~pc -- 0.03~pc scales that \citet{downes2023} find are also dynamically important.

%=======================================================================
\section{Conclusions}
%=======================================================================

In this paper, we present maps of the CO, \tco, and \cdo\ emission from the \mm, a large cloud complex with multiple dust pillars located in the heart of the Carina Nebula. 
A dendrogram analysis reveals a coherent, connected structure with 
three individual pillars. 
We detect eight continuum cores within the \mm. 
Most continuum cores are associated with a \cdo\ clump, but not all \cdo\ clumps have a continuum counterpart. 
The rarer isotopologues DCN J=3-2 and $^{13}$CS J=5-4 are detected in two clumps located in the region with the highest column density.

The \mm\ region experiences an order of magnitude higher ionizing flux from the nearby star clusters than the flux incident on other dust pillars in Carina observed with similar tracers and angular resolution. 
However, bulk pillar properties like 
the average velocity dispersion derived from moment~2 maps are similar for all pillars, regardless of their irradiation. 

A more detailed analysis to isolate turbulent motions reveals a turbulent driving parameter, $b=0.55$, consistent with compressive turbulence dominating in the \mm. 
The derived $b$ is within the range of values found by \citet{menon2021} for the pillars from \citet{klaassen2020}.  
The derived Mach number for the \mm\ is a factor of 2 higher than that found for other pillars in the Carina region, either reflecting a stronger shock from the more intense UV field, or, more likely, is artificially inflated by the broad velocity distribution of the \mm.

The similarity of pillar properties across a range of incident ionizing fluxes contrasts with studies of other irradiated clouds in the Carina Nebula. 
\citet{rebolledo2020} compared two regions with an order of magnitude difference in the incident radiation and find evidence for more turbulence and fewer (but higher mass) cores in the more heavily irradiated cloud. 
From a different analysis and dataset of the same region presented by \citet{rebolledo2020}, 
\citet{downes2023} argue that differences between the pillar results and the cloud results are not inconsistent because pillar kinematics may reflect dynamical compression. 
We argue that this may be true if the different morphologies observed today result from different initial densities that aided or prevented UV irradiation from compressing the local cloud into a pillar. 
Pre-existing overdensities may also explain the observed difference in star-forming activity. 
Cores in the Western Wall / North Cloud appear starless whereas a fraction of the cores in the \mm\ and other pillars drive prominent jets, signifying a more advanced evolutionary stage \citep[i.e.\ HH~901 and HH~902 in the \mm; see][]{cortes-rangel2020}.

Future work probing a broader range of environments with the angular resolution to probe both the large scales of turbulence driving and the small scales where its consequences are most evident will help resolve this tension.

\section*{Acknowledgements}
We would like to thank the referee, Neal Evans, for a prompt and thoughtful report that improved the manuscript. 
We would like to thank Shyam Menon for a careful reading of the manuscript and thoughtful feedback. 
% MRR
MR was partially supported by an ESO fellowship. 
% DI 
DI was funded by the European Research Council (ERC) via the ERC Synergy Grant ECOGAL (grant 855130).
% ALMA
This paper makes use of the
following ALMA data: ADS/JAO.ALMA\#2018.1.01001.S.
ALMA is a partnership of ESO (representing its member
states), NSF (USA) and NINS (Japan), together with
NRC (Canada) and NSC and ASIAA (Taiwan) and KASI
(Republic of Korea), in cooperation with the Republic of
Chile. The Joint ALMA Observatory is operated by ESO,
AUI/NRAO and NAOJ. 
The National Radio Astronomy Observatory is a facility of the National Science Foundation operated under cooperative agreement by Associated Universities, Inc.
% Hubble 
This work uses observations made
with the NASA/ESA Hubble Space Telescope, obtained
from the Data Archive at the Space Telescope Science Institute,
which is operated by the Association of Universities
for Research in Astronomy, Inc., under NASA contract
NAS 5-26555. The HST observations are associated with GO-12050. 
This research made use of Astropy\footnote{http://www.astropy.org}, a community developed core Python package for Astronomy 
\citep{astropy:2013,astropy:2018}. This research
made use of APLpy, an open-source plotting package
for Python \citep{robitaille2012}.
This research made use of the following software packages: 
astrodendro, a Python package to compute dendrograms of Astronomical data (http://www.dendrograms.org/); SCIMES, a Python package to find relevant structures into dendrograms of molecular gas emission using the spectral clustering approach \citep{colombo2015}; 
and 
TurbuStat, a Python package to compute 14 turbulence-based statistics described in the astronomical literature \citep{koch2017,koch2019}.

%%%%%%%%%%%%%%%%%%%%%%%%%%%%%%%%%%%%%%%%%%%%%%%%%%
\section*{Data Availability}

%The inclusion of a Data Availability Statement is a requirement for articles published in MNRAS. Data Availability Statements provide a standardised format for readers to understand the availability of data underlying the research results described in the article. The statement may refer to original data generated in the course of the study or to third-party data analysed in the article. The statement should describe and provide means of access, where possible, by linking to the data or providing the required accession numbers for the relevant databases or DOIs.

The ALMA data used in this study are publicly available from the ALMA archive \footnote{\href{https://almascience.nrao.edu/aq/?result_view=observations}{https://almascience.nrao.edu/aq/?result\_view=observations}} under the program ID number ADS/JAO.ALMA\#2018.1.01001.S. 
Data from \emph{HST} are publicly available via the MAST archive\footnote{\href{https://mast.stsci.edu/portal/Mashup/Clients/Mast/Portal.html}{https://mast.stsci.edu/portal/Mashup/Clients/Mast/Portal.html}}.



\bibliographystyle{mnras}
\bibliography{bibliography_mrr} % if your bibtex file is called example.bib

%\begin{thebibliography}{99}
%\end{thebibliography}

%%%%%%%%%%%%%%%%%%%%%%%%%%%%%%%%%%%%%%%%%%%%%%%%%%

%%%%%%%%%%%%%%%%% APPENDICES %%%%%%%%%%%%%%%%%%%%%

\appendix

\section{Optical depth maps}\label{appendix:tau_maps}

\begin{figure*}
	\includegraphics[width=0.45\textwidth]{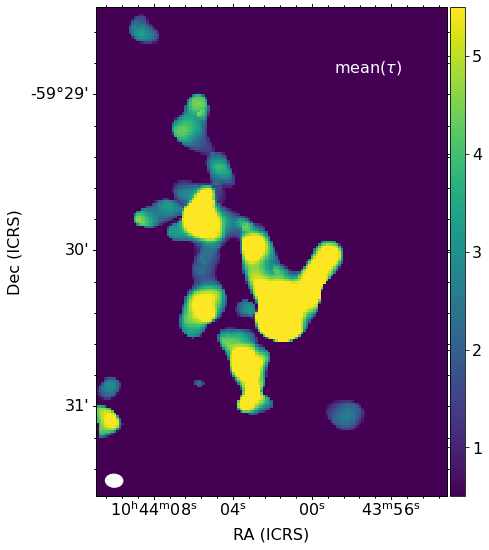}
	\includegraphics[width=0.45\textwidth]{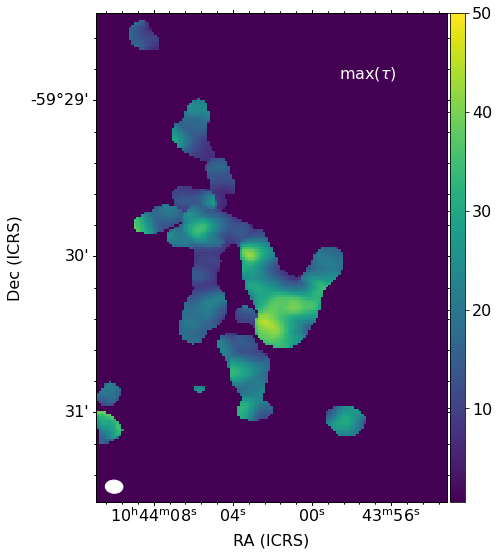}
    \includegraphics[width=0.45\textwidth]{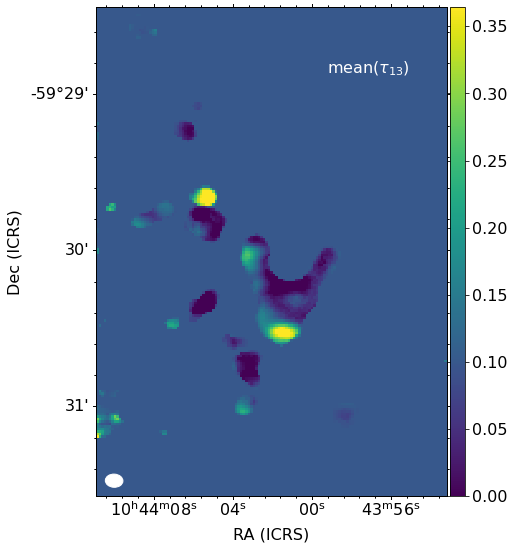}
	\includegraphics[width=0.45\textwidth]{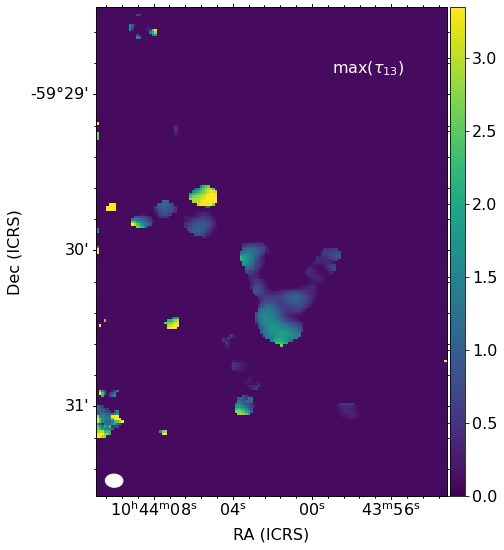}
    \caption{
    \textbf{Left:} The mean and 
    \textbf{Right:} maximum $\tau$ of CO (top) and \tco\ (bottom) over the velocity range $-10.5 < v < -2$~\kms.  }
    \label{fig:tau_map}
\end{figure*}

Maps of the average and maximum $^{12}$CO optical depth are shown in Figure~\ref{fig:tau_map}. 
The median value for the \mm\ and each individual pillar are reported in Table~\ref{t:molecular_props}. 

\section{Column density maps}\label{appendix:Ncol_maps}

\begin{figure*}
	\includegraphics[scale=0.325]{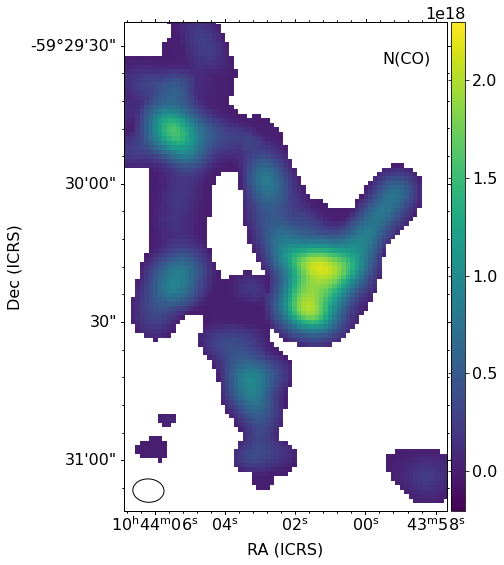}
	\includegraphics[scale=0.325]{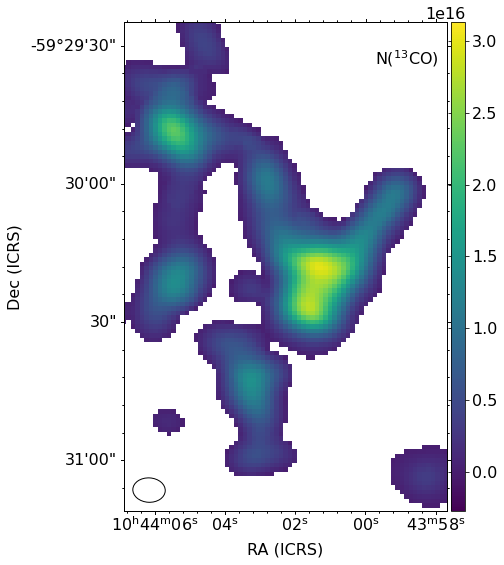}
	\includegraphics[scale=0.325]{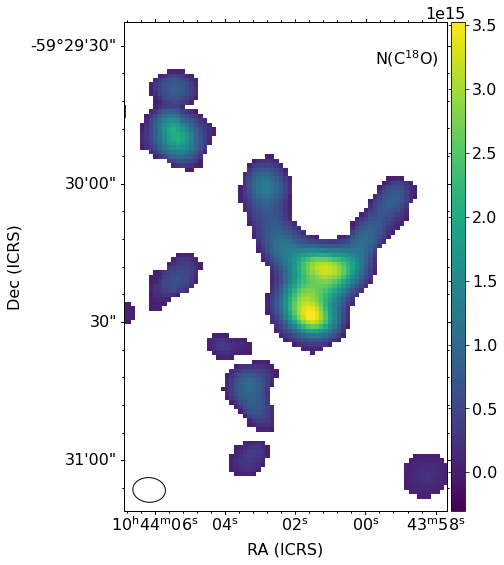}
    \caption{
    Column density maps of the CO isotopologues over the velocity range $-10.5 < v < -2$~\kms.  }
    \label{fig:Ncol_maps}
\end{figure*}

The spatially-resolved, optical-depth corrected column density of each of the three CO isotopologues is shown in Figure~\ref{fig:Ncol_maps}. 
Peak and median values for the \mm\ and each individual pillar are reported in Table~\ref{t:molecular_props}. 

\section{Turbulence Driving Parameter analysis for the individual pillars}\label{appendix:pillar_vturb}

We repeat the analysis in Section~\ref{ss:turb_motions} to isolate turbulent motions in each pillar of the \mm\ complex individually. 
The pillars in the \mm\ are all at slightly different velocities and these differences are not cleanly removed with a linear gradient. 
These real velocity differences therefore inflate any measure of the velocity dispersion for the \mm\ as a whole. Ideally, each pillar would be analyzed separately to remove these systematic motions.

\begin{figure*}
    \centering
    \includegraphics[width=0.355\textwidth,trim=0mm -10mm 0mm 0mm]{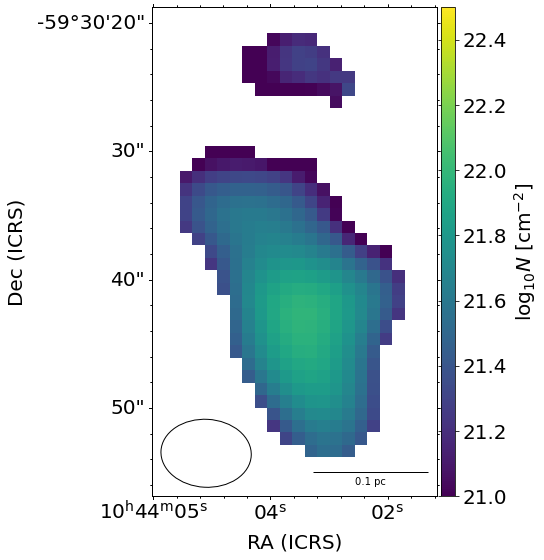}
    \includegraphics[width=0.605\textwidth]{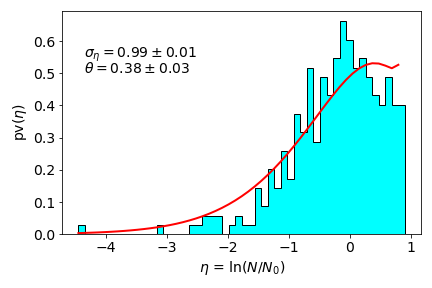}
    \includegraphics[width=0.355\textwidth,trim=0mm -10mm 0mm 0mm]{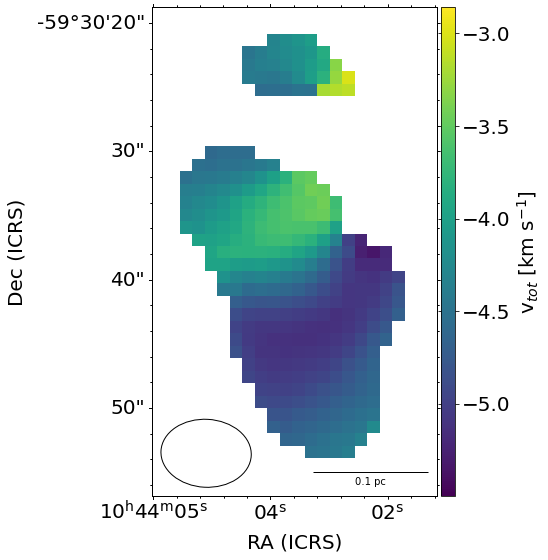}
    \includegraphics[width=0.605\textwidth]{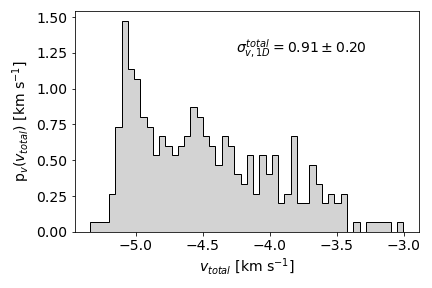}
    \includegraphics[width=0.355\textwidth,trim=0mm -10mm 0mm 0mm]{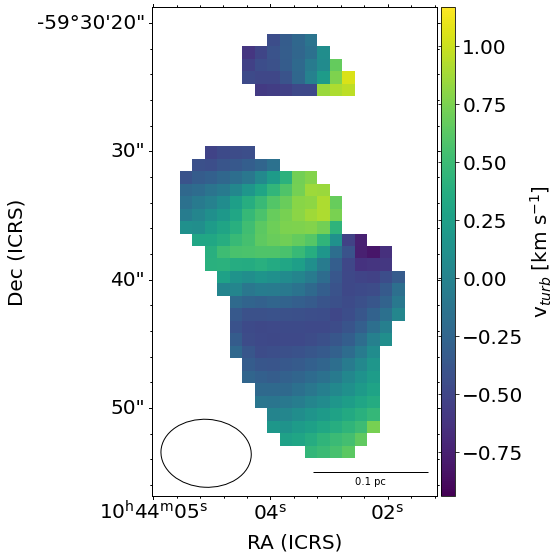}
    \includegraphics[width=0.605\textwidth]{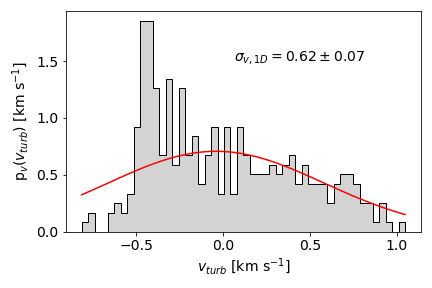}
    \caption{Same as Figure~\ref{fig:turb_derv} but for the HH~901~pillar alone. }
    \label{fig:hh901_turb_derv}
\end{figure*}

\begin{figure*}
    \centering
    \includegraphics[trim=0mm -20mm 0mm 0mm,width=0.405\textwidth]{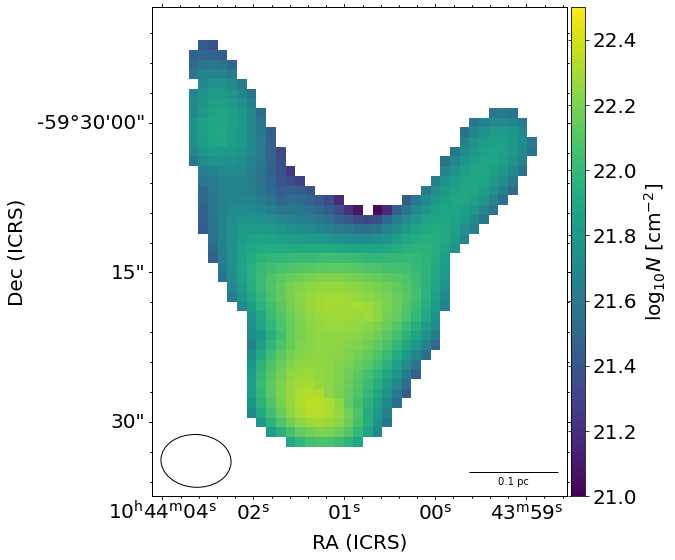}
    \includegraphics[width=0.585\textwidth]{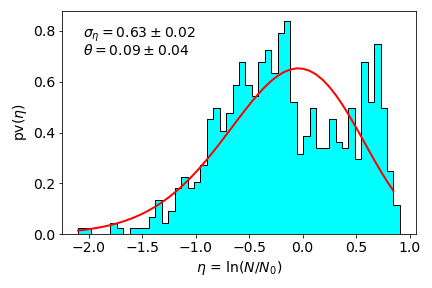}
    \includegraphics[trim=0mm -20mm 0mm 0mm,width=0.405\textwidth]{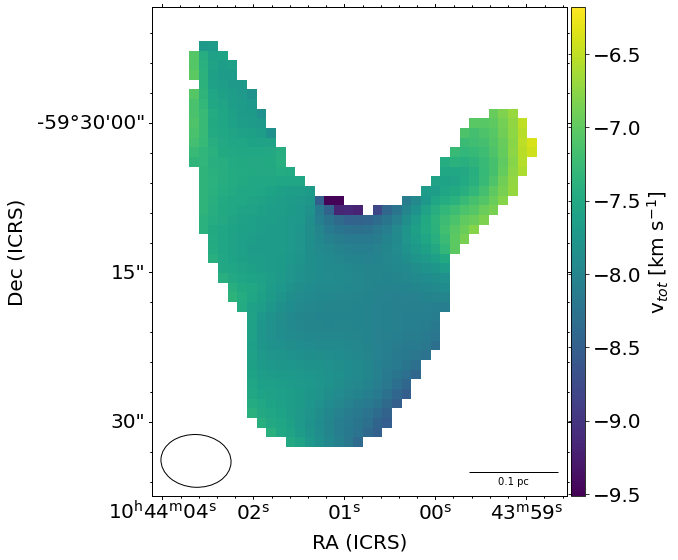}
    \includegraphics[width=0.585\textwidth]{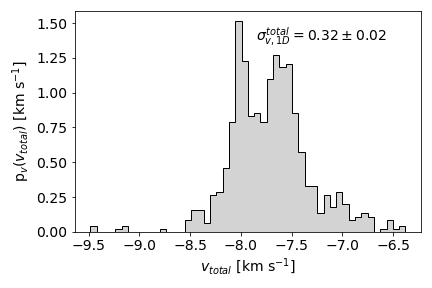}
    \includegraphics[trim=0mm -20mm 0mm 0mm,width=0.405\textwidth]{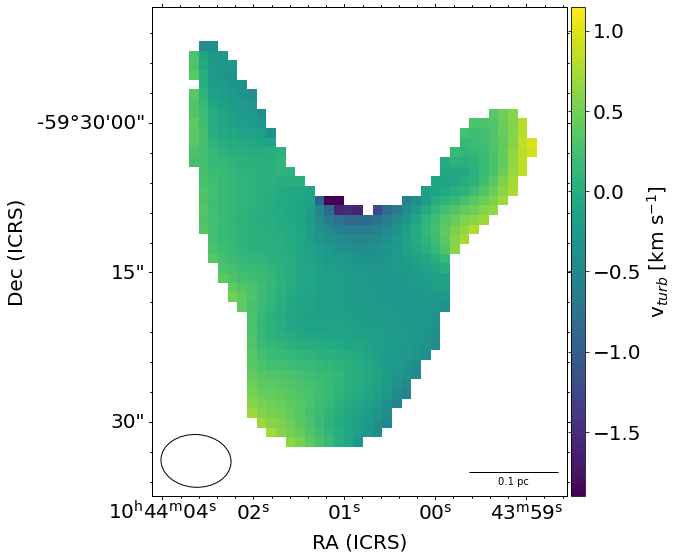}
    \includegraphics[width=0.585\textwidth]{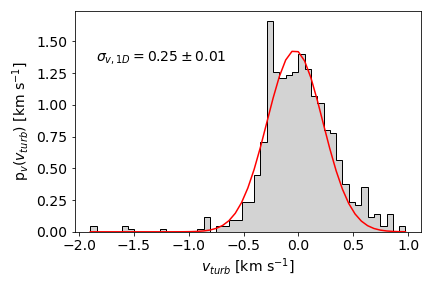}
    \caption{Same as Figure~\ref{fig:turb_derv} but for the HH~902~pillar alone.}
    \label{fig:hh902_turb_derv}
\end{figure*}

\begin{figure*}
    \centering
    \includegraphics[width=0.345\textwidth,trim=0mm -15mm 0mm 0mm]{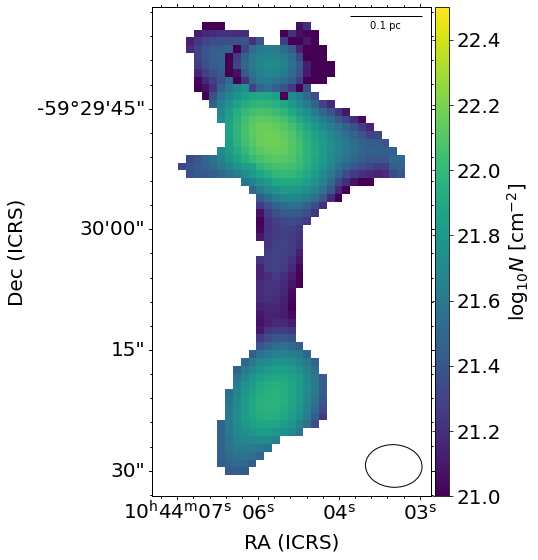}
    \includegraphics[width=0.605\textwidth]{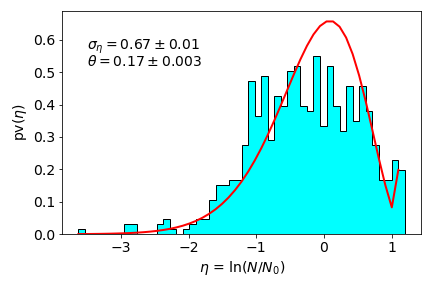}
    \includegraphics[width=0.35\textwidth,trim=0mm -15mm 0mm 0mm]{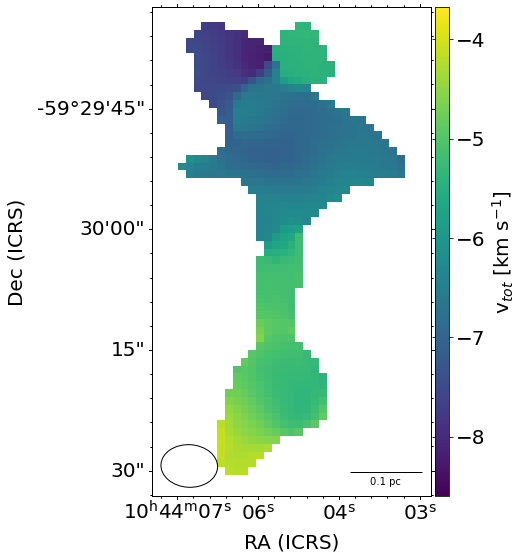}
    \includegraphics[width=0.605\textwidth]{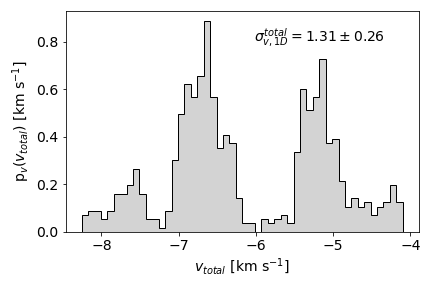}
    \includegraphics[width=0.35\textwidth,trim=0mm -10mm 0mm 0mm]{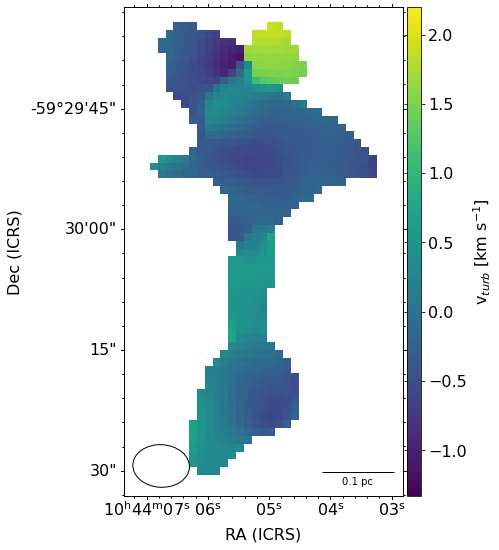}
    \includegraphics[width=0.605\textwidth]{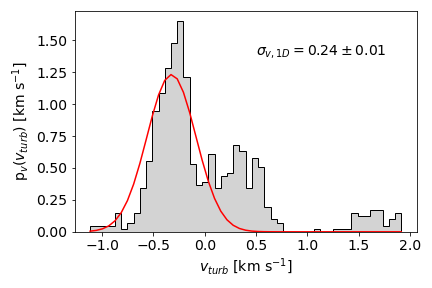}
    \caption{Same as Figure~\ref{fig:turb_derv} but for the HH~1066~pillar alone.}
    \label{fig:hh1066_turb_derv}
\end{figure*}

Figures~\ref{fig:hh901_turb_derv}, \ref{fig:hh902_turb_derv}, and \ref{fig:hh1066_turb_derv} show the results of this analysis for the individual pillars and Table~\ref{t:indiv_turb_params} reports the derived values with 1$\sigma$ uncertainties. 
We find unphysical values for many parameters, including the turbulent driving parameter $b$. 
This is almost certainly a reflection of the low resolution of our data compared to the size of the pillars: the width of the HH~901 and HH~1066 pillars are both comparable to the size of the beam.

To test this, we degraded the resolution of the data to 1/4 of the native resolution \citep[as in][]{sharda2018}. 
Repeating the analysis, we find that $\sigma_{v,3D}$ changes by $>$20\% for the separate pillars. 
For the  \mm\ as a whole, derived values change by $\sim$6\% using the lower-resolution data. 
Formal error propagation does not capture this underlying problem, leading to modest uncertainties on unphysical values in Table~\ref{t:indiv_turb_params}. 
A more reliable analysis requires 
higher angular observations that better resolve the individual pillars.

%%--------------------------------------------------
% HH 901
%%--------------------------------------------------

%%--------------------------------------------------
% HH 902 
%%--------------------------------------------------

%%--------------------------------------------------
% HH 1066
%%--------------------------------------------------

%%----------------------------------------------------------------
\begin{table}
\caption{Turbulence parameters each pillar in the \mm\ individually. \label{t:indiv_turb_params}}
%\centering
\begin{footnotesize}
\begin{tabular}{cccc}
\hline
Pillar & HH~901 & HH~902 & HH~1066 \\ 
\hline
A [pc$^2$] & 0.07$\pm$0.007 & 0.18$\pm$0.02 & 0.20$\pm$0.02 \\
$N_0$ [10$^{21}$ cm$^{-2}$] & 3.09$^{+3.5}_{-1.6}$ & 7.59$^{+6.0}_{-3.4}$ & 3.47$^{+4.1}_{-1.9}$ \\
$M$ [\Msun] & 3.31$\pm$2.64 & 17.2$\pm$13.2 & 8.13$\pm$7.51 \\
n [$10^3$cm$^{-3}$] & 4.16$\pm$1.06 & 6.10$\pm$0.76 & 2.95$\pm$0.80 \\
$\sigma^{M2}_{v,3D}$ [\kms] & 0.31$\pm$0.23 & 0.27$\pm$0.15 & 0.26$\pm$0.26 \\
$\alpha_{vir}$ & 16.3$\pm$1.90 & 0.76$\pm$0.12 & 1.24$\pm$0.33 \\
t$_{\mathrm{ff}}$/t$_{\mathrm{turb}}$ & 1.73$\pm$0.10 & 0.37$\pm$0.03 & 0.48$\pm$0.06 \\
$\sigma_{\eta}$ & 0.99$\pm 0.01$ & 0.63$\pm 0.02$ & 0.67$\pm 0.01$ \\
$\sigma_{N/N_0}$ & 0.52$\pm 0.01$ & 0.46$\pm 0.02$ & 0.45$\pm 0.01$ \\
$\mathcal{R}^{1/2}$ & 0.09$\pm0.003$ & 0.13$\pm0.0004$ & 0.09$\pm0.001$ \\
$\sigma_{\rho/\rho_0}$ & 5.76$\pm$0.21 & 3.49$\pm$.15 & 4.75$\pm$.08 \\
$\sigma^{\mathrm{total}}_{v,1D}$ [\kms] & 0.91$\pm$0.20 & 0.32$\pm$0.02 & 1.31$\pm$0.26 \\
$\sigma_{v,1D}$ [\kms] & 0.62$\pm$0.07 & 0.25$\pm$0.01 & 0.24$\pm$0.01 \\
$\sigma_{v,3D}$ [\kms] & 1.07$\pm$0.13 & 0.44$\pm$0.02 & 0.41$\pm$0.02 \\
$\mathcal{M}$ & 3.58$\pm$0.43 & 1.47$\pm$0.05 & 1.38$\pm$0.07 \\
$b$ & 1.61$\pm$0.2 & 2.37$\pm$0.13 & 3.45$\pm$0.19 \\
\hline 
\end{tabular}
\end{footnotesize}
\end{table}
%%----------------------------------------------------------------

%%%%%%%%%%%%%%%%%%%%%%%%%%%%%%%%%%%%%%%%%%%%%%%%%%

% Don't change these lines
\bsp	% typesetting comment
\label{lastpage}
\end{document}